\documentclass[12pt]{article}
\usepackage{epsf}

\makeatletter
\@addtoreset{equation}{section}

\renewcommand{\thefootnote}{\fnsymbol{footnote}}
\makeatother

\begin{document}

\thispagestyle{empty}
%
\begin{flushright}
TIT/HEP--468 \\
OU-HET 391\\
{\tt hep-th/0108179} \\
August, 2001 \\
\end{flushright}
\vspace{3mm}
\begin{center}
{\Large
{\bf BPS Walls and Junctions} \\
\vspace{2mm}
{\bf in SUSY Nonlinear Sigma Models
}} 
\\[12mm]
%
\vspace{5mm}

\normalsize

  {\large \bf 
  Masashi~Naganuma~$^{a}$}
\footnote{\it  e-mail address: 
naganuma@th.phys.titech.ac.jp
},  
 {\large \bf 
Muneto~Nitta~$^{b}$}
\footnote{\it  e-mail address: 
nitta@het.phys.sci.osaka-u.ac.jp}\footnote{
Address after September 1, Department of Physics, Purdue University, 
West Lafayette, IN 47907-1396, USA.},
~and~~  {\large \bf 
Norisuke~Sakai~$^{a}$}
\footnote{\it  e-mail address: 
nsakai@th.phys.titech.ac.jp} 

\vskip 1.5em

{ \it $^{a}$Department of Physics, Tokyo Institute of 
Technology \\
Tokyo 152-8551, JAPAN  \\
and \\
  $^{b}$Department of Physics, Osaka University 
560-0043, JAPAN }
\vspace{5mm}
{\bf Abstract}\\[5mm]
{\parbox{13cm}{\hspace{5mm}
BPS walls and junctions are studied in ${\cal N}=1$ 
SUSY nonlinear sigma models in four spacetime dimensions. 
New BPS junction solutions connecting $N$ discrete vacua 
are found 
for nonlinear sigma models with several chiral scalar superfields. 
A nonlinear sigma model with a single chiral scalar superfield 
is also found which has a moduli space of the topology of $S^1$ 
and admits BPS walls and junctions connecting arbitrary points in moduli 
space. 
SUSY condition in nonlinear sigma models are classified either as 
stationary points of superpotential or singularities of the K\"ahler metric 
in field space. 
The total number of SUSY vacua is invariant under holomorphic field 
redefinitions if we count ``runaway vacua'' also.  
}}
\end{center}
\vfill
\newpage
\setcounter{page}{1}
\setcounter{footnote}{0}
\renewcommand{\thefootnote}{\arabic{footnote}}

\section{Introduction}\label{INTRO}
\vspace{5mm}

Supersymmetric theories stand as one of the most attractive and well-studied 
theories to build unified theories beyond the standard model \cite{DGSWR}. 
More recently models with extra dimensions have been studied extensively 
\cite{LED}, \cite{RS}. 
Thier idea is called brane-world scenario, since our world is assumed 
to be realized on an extended topological defects 
such as domain walls or various branes. 
Supersymmetry (SUSY) can also be implemented in these models 
and helps the construction of the extended objects. 
SUSY breaking mechanisms have been discussed in the context of 
brane world scenario \cite{DvaliShifman}--\cite{MSSS2}. 
Preservation of part of the SUSY gives 
the so-called BPS states \cite{WittenOlive}, which have been very useful 
in analyzing various nonperturbative effects. 
The coexistence of BPS walls preserving orthogonal combinations of 
SUSY gives a non-BPS state which provides a new mechanism of SUSY 
breaking \cite{MSSS}.
Domain walls typically conserve half of the SUSY and are called 
${1 \over 2}$ BPS states \cite{CGR}, \cite{DW}. 
The junction of these domain walls 
have also been studied \cite{AbrahamTownsend}--\cite{GrSh} and 
preserves a quarter of original SUSY. 
They should be useful to consider brane-world scenario based on theories 
with higher dimensions and more SUSY. 
More recently, we have succeeded to construct an analytic solution 
for the junction in the ${\cal N}=1$ SUSY field theories 
in four dimensions \cite{OINS}. 
Our exact solution has a $Z_3$ symmetry and has given several unexpected 
informations such as the sign of the central charge, and non-normalizability 
of the Nambu-Goldstone fermion modes \cite{INOS}. 

In order to consider models with extra dimensions, 
we need to discuss 
supersymmetric theories in spacetime with dimensions higher than four. 
They should have at least eight supercharges. 
These SUSY are so restirictive that possible potential terms are 
severely constrained. 
The only nontrivial interactions come either from nonlinearity of 
kinetic term (nonlinear sigma model) or gauge interactions 
 \cite{AlvarezFreedman}--\cite{SierraTownsend}. 
If reduced to four dimensions, the theory has at least ${\cal N}=2$ SUSY. 
It has been shown that one has to consider nonlinear sigma model with 
a nontrivial K\"ahler metric in field space if one wants to have 
nontrivial interactions with only hypermultiplets in ${\cal N}=2$ 
theories in four-dimensions. 
The only possible potential term is given by the square of a tri-holomorphic 
Killing vector field of the (hyper-)K\"ahler metric and the vacua arise as 
 singularities of the (hyper-)K\"ahler metric instead of stationary 
points of superpotential \cite{AlvarezFreedman}, \cite{GPT}. 
This potential term can also be understood as due to a dimensional reduction 
with nontrivial twists similarly to the Sherk-Schwarz mechanism 
\cite{ScherkSchwarz}. 
Therefore we need to consider the nonlinear sigma model for 
such SUSY theories if we wish to obtain an interesting solutions like 
domain walls and/or junctions using only hypermultiplets \cite{GPT}, 
\cite{NNS1}. 

The purpose of our paper is to study the nonlinear sigma model 
in a simpler context of ${\cal N}=1$ SUSY theories in four dimenions 
to obtain walls and/or junctions as BPS solutions. 
This study should be useful in its own right, and will serve as a starting 
point for a more difficult case of larger number of SUSY charges. 
We find a nonlinear sigma model with several chiral scalar superfields 
which admits a new exact junction solutions connecting $N$ discrete 
vacua.  
The model and the solution are generalizations of our original 
 $Z_3$ symmetric vacua to a generic $N$ discrete vacua.  
 Another nonlinear sigma model with a single chiral 
scalar field is also obtained which admits our $Z_3$ symmetric 
junction solution as an exact BPS solution. 
We find that this single field model has a moduli space with $S^1$ topology 
and that it admits BPS walls 
and junctions connecting arbitrary points in the moduli space. 

We examine the SUSY condition in the case of the nonlinear sigma model and 
find that the SUSY vacua can come 
from singularities of the K\"ahler metric in field space similarly to 
the ${\cal N}=2$ nonlinear sigma model. 
In the ${\cal N}=1$ model, the SUSY vacua can also come from stationary 
points of superpotential as in the linear sigma model. 
We also find that the field redefinition can turn these stationary points 
into sigularities and vice versa, but it preserves the number and character 
of the SUSY vacua if those at infinity in field space are included. 
We identify those nonlinear sigma models which can be mapped into 
linear sigma models and call them holomorphically factorizable. 
Even in such models, the nonlinear sigma models are 
sometimes more useful by revealing SUSY vacua which are usually 
ignored as ``runaway vacua'' at infinity. 
We also find that choosing superpotential itself as one of the chiral scalar 
superfield is quite useful and sometimes natural in discussing the 
BPS walls and junctions, since the domain wall 
configuration becomes a straight line in the complex superpotential plane. 
We deal with classical field theories in this paper and 
will postpone to discuss questions on quantization and quantum 
effects for subsequent studies. 

In sect.2, the condition for SUSY vacua is established in nonlinear sigma 
models. 
In sect.3, BPS equation is reviewed and the 
the choice of superpotential as a chiral scalar superfield is advocated 
to study BPS states in nonlinear sigma models. 
In sect.4, field redefinition ambiguities are studied and 
usefulness of the nonlinear sigma model in revealing a ``runaway vacuum'' 
as a legitemate vacuum is illustrated in a simple model. 
In sect.5, a nonlinear sigma model with a single chiral scalar superfield 
is worked out which admits our $Z_3$-symmetric junction as a BPS solution. 
Walls and junctions connecting arbitrary points in moduli space are 
also constructed. 
Sect.6 is devoted to constructing a nonlinear sigma model with $N$ discrete 
vacua which admits a BPS junction solution as a generalization of 
our $Z_3$-symmetric junction. 
We work out up to $N=4$ case explicitly. 

\vspace{5mm}
\section{BPS equations 
in nonlinear sigma models
}\label{sc:BPS_eq}
\subsection{SUSY vacua 
in nonlinear sigma models
}\label{sc:VACUA}

We shall examine the condition of supersymmetric 
vacuum 
in the case of general ${\cal N}=1$ nonlinear sigma model 
with an arbitrary superpotential ${\cal W}$ in four spacetime 
dimensions. 
The chiral scalar superfields and the K\"ahler 
potential for the kinetic term are denoted as $\Phi^i$ and 
$K\left(\Phi, \Phi^*\right)$, respectively. 
 Following the convention in Ref.\cite{WessBagger}, the Lagrangian is 
 given by 
\begin{eqnarray}
{\cal L}
&\!=&\!\int d^2\theta d^2\bar{\theta} K(\Phi,\Phi^{* }) 
 + \left[ \int d^2\theta {\cal W}(\Phi) 
+ \mbox{h.c.} \right],
\nonumber \\
&\!=&\! 
 K_{i j^*}(A, A^*)\left(-\partial_\mu A^{*j} \partial^\mu A^i + 
F^{*j}F^i 
+{i \over 2} \partial_\mu \bar \psi^j \bar \sigma^\mu \psi^i 
-{i \over 2} \bar \psi^j \bar \sigma^\mu \partial_\mu \psi^i 
\right)
\nonumber \\
&\!+&\! F^j {\partial {\cal W} \over \partial A^j} 
-{1 \over 2} \psi^i \psi^j 
{\partial {\cal W} \over \partial A^i \partial A^j} 
+ F^{*j} {\partial {\cal W}^* \over \partial A^{*j}} 
-{1 \over 2} \bar \psi^i \bar \psi^j 
{\partial {\cal W}^* \over \partial A^{*i} \partial A^{*j}} .
\label{generalNLSM_auxiliary}
\end{eqnarray}
where $ K_{i j^*}=\partial^2 K(A^*, A) /\partial A^i\partial 
A^{*j}$ 
is the K\"ahler metric. 

The equations of motion for auxilary fields $F^i$ are given by 
\begin{equation}
K_{i j^*} F^{*j} + {\partial {\cal W} \over \partial A^{i}} = 0 .
\end{equation}
After eliminating the auxiliary fields $F^i$ the Lagrangian 
becomes  
\begin{eqnarray}
{\cal L}
&\!\!\!=&\!\!\! 
 K_{i j^*}(A, A^*)\left(-\partial_\mu A^{*j} \partial^\mu A^i 
+{i \over 2} \partial_\mu \bar \psi^j \bar \sigma^\mu \psi^i 
-{i \over 2} \bar \psi^j \bar \sigma^\mu \partial_\mu \psi^i 
\right)
\nonumber \\
&\!\!\!
&\!\!\!
-{1 \over 2} \psi^i \psi^j 
{\partial {\cal W} \over \partial A^i \partial A^j} 
-{1 \over 2} \bar \psi^i \bar \psi^j 
{\partial {\cal W}^* \over \partial A^{*i} \partial A^{*j}} 
- V (A, A^*) . 
\end{eqnarray}
The scalar potential $V (A, A^*) $ is given by 
\begin{equation}
V (A, A^*) = K^{i j^*} {\partial {\cal W} \over \partial A^{i}} 
{\partial {\cal W}^* \over \partial A^{*j}} 
=
K_{i j^*} F^{i}  F^{*j} 
\label{eq:scalarpotential}
\end{equation}
where $ K^{i j^*} $ is the inverse of the K\"ahler metric  $ 
K_{i j^*}$. 

In order to respect the holomorphy, 
field redefinitions of nonlinear sigma models must be 
restricted to holomorphic redefinitions of chiral scalar superfields 
in the case of ${\cal N}=1$ SUSY theories. 
By field redefinitions, 
various quantities such as component fields 
and K\"ahler metric transform covariantly 
\begin{eqnarray}
A^i 
&\!\!\!\rightarrow &\!\!\!
A'^i =  A'^i\left( A \right), 
\nonumber \\
\psi^{i} 
&\!\!\!\rightarrow &\!\!\!
\psi'^i = {\partial A'^i \over \partial A^{j}} \psi^{j}, 
\nonumber \\
F^{i} 
&\!\!\!\rightarrow &\!\!\!
F'^i = {\partial A'^i \over \partial A^{j}} F^j 
- \frac{1}{2}
{\partial^2 A'^i \over \partial A^{j}\partial A^k} \psi^{j} \psi^k. 
\label{eq:fieldredefinition}
\end{eqnarray}
\begin{equation}
K_{i j^*}  \rightarrow K'_{i j^*} 
= {\partial A'^{k} \over \partial A^{i}} 
 {\partial A'^{*l} \over \partial A^{*j}} 
K'_{k l^*} .
\label{eq:transformation_kahler}
\end{equation}
where $A'^i (A)$ denotes an arbitrary function of the scalar field $A$ 
to define the redefinition of superfields $\Phi^i$. 
On the other hand, the K\"ahler potential $K (A, A^*)$ 
and the scalar potential $V (A, A^*)$ are 
invariant under the field redefinitions 
\begin{eqnarray}
K (A, A^*)
 &\!\! 
\rightarrow 
 &\!\! 
K (A', A'^*)\equiv K \left(A(A'), A^*(A'*)\right), 
\nonumber \\
V (A, A^*) 
 &\!\! 
\rightarrow 
 &\!\! 
V (A', A'^*)\equiv V \left(A(A'), A^*(A'*)\right) . 
\label{eq:invariance_sclarpotential}
\end{eqnarray}

The condition of SUSY vacuum is given by the vanishing 
vacuum energy 
density 
\begin{equation}
0=V (A, A^*) = 
K^{i j^*} {\partial {\cal W} \over \partial A^{i}} 
{\partial {\cal W}^* \over \partial A^{*j}} 
=
K_{i j^*} F^{i}  F^{*j} . 
\label{eq:SUSYcondition}
\end{equation}

To simplify matters, let us take the case of the nonlinear 
sigma model 
with only a single chiral scalar superfield $\Phi$. 
We find that there are two cases for 
the SUSY vacuum in the nonlinear sigma model :
\begin{enumerate}
\item
Stationary point of superpotential which is not a zero 
of the K\"ahler metric 
\begin{equation}
{\partial {\cal W} \over \partial A} = 0, 
\quad  {\rm and} \quad 
K_{A A^*} \not= 0 
\quad \left( \; K^{A A^*} \not= \infty \; \right) , 
\label{eq:stationary_superpotential}
\end{equation}

\item
Singularity of the the K\"ahler metric which is not a 
singularity 
of the derivative of the superpotential
\begin{equation}
K_{A A^*} = \infty 
\quad \left( \; K^{A A^*} = 0 \; \right), 
\quad {\rm and} \quad 
{\partial {\cal W} \over \partial A} \not= \infty . 
\label{eq:singularity_kahlermetric}
\end{equation}
 
\end{enumerate}

It is interesting to observe that the vanishing $F$ term ($F^i=0$) is 
not necessary nor sufficient for the unbroken SUSY in nonlinear sigma models, 
since the K\"ahler metric in Eq.(\ref{eq:SUSYcondition}) can have zeros or 
singularities. 
The holomorphic field redefinition (\ref{eq:fieldredefinition}) 
can transform a stationary point of the superpotential into 
a singularity of the K\"ahler metric and vice versa. 
However, the total number 
of SUSY vacua 
is conserved, since the scalar potential in Eq.(\ref{eq:SUSYcondition}) is 
invariant under field redefinitions contrary to $F$ terms. 
Therefore no new SUSY vacua can appear or disappear 
by the field redefinitions in generic circumstances. 
However, there is an exceptional situation where new SUSY vacua 
can properly be recognized only by using the field redefinition. 
Suppose that there is a SUSY vacuum at infinity in the field 
space. 
In such a situation, the SUSY vacuum at infinity is 
called a runaway vacuum and is usually discarded from the 
list of 
SUSY vacua. 
It is often possible to make a field redefinition to bring the 
runaway vacuum at infinity to a finite point in field space. 
Then we recognize it as one of the legitimate SUSY vacua 
rather than 
a runaway vacuum. 
This is the exceptional situation where a new legitimate 
SUSY vacuum arises from a hidden runaway vacuum. 
It is also possible to have the reversed situation where 
SUSY vacuum disappears as a runaway vacuum by a field 
redefinition. 
We shall illustrate this phenomenon in a simple model later.

\vspace{5mm}

\subsection{BPS equation and superpotential as a field}
\label{sc:BPSeq}
In order to consider the junction configuration later, we need 
to consider 
a field configuration which is nontrivial in two-dimensional 
space. 
Without loss of generality, we can assume that 
the field configuration depends on $x^1$ and $x^2$ only. 
By requiring the conservation of one supercharge out of four, 
we find that there are two possible BPS equations for the 
${1 \over 4}$ BPS state \cite{OINS}--\cite{INOS}. 
The first choice is given by 
\begin{equation}
2 
{\partial A^i \over \partial z} 
=- \Omega_{-} F^i 
= \Omega_{-}K^{ i j^*} \frac{\partial {\cal W}^*}{ \partial A^{*j}}, 
\quad 
\Omega_{-}\equiv i\frac{
 -i Z_1^*-Z_2^* 
}
{|
 -i Z_1^*-Z_2^* 
 |} .
\label{Be2}
\end{equation}
where $z=x^1+i x^2, \bar{z}=x^1-i x^2$ are complex 
coordinates, 
and the constant phase factor $\Omega_{-}$ is given by the 
central charges $Z_1$ and $Z_2$. 
The second choice of ${1 \over 4}$ BPS equations 
corresponds to the 
conservation of another orthogonal linear combination of 
supercharges 
and is given by 
\begin{equation}
2 
{\partial A^i \over \partial \bar{z}} 
=- \Omega_{+} F^i 
= \Omega_{+}K^{ i j^*} \frac{\partial {\cal W}^*}{ \partial A^{*j}}, 
\quad 
\Omega_{+}\equiv i \frac{
 -i Z_1^*+Z_2^* 
 }
{|
 -i Z_1^*+Z_2^* 
 |},
\label{Be1}
\end{equation}
with a similar constant phase factor $\Omega_{+}$. 
This second BPS equation is sometimes called the anti-BPS equation. 
If both ${1 \over 4}$ BPS equations (\ref{Be2}) and (\ref{Be1}) 
are satisfied at the same time, we obtain an ${1 \over 2}$ 
BPS state. 

Multiplying the BPS equation (\ref{Be2}) by 
$\partial {\cal W}/ \partial A^{i}$, one finds 
\begin{equation}
2 
{\partial {\cal W} \over \partial z} =
2 {\partial {\cal W} \over \partial A^i}
{\partial A^i \over \partial z} 
= \Omega_{-}K^{ i j^*} 
 {\partial {\cal W} \over \partial A^i}
\frac{\partial {\cal W}^*}{ \partial A^{*j}}.
\label{eq:BPSin_superpotential}
\end{equation}
If we consider a one-dimensional configuration which depends on only one 
linear combination of coordinates as in the case of the domain wall, 
we have a simple theorem 
\cite{CGR}, \cite{CHT}--\cite{INOS}: 
the configuration becomes a straight line 
if we consider it in the space of superpotential. 
To show this, let us assume that the configuration depends only on 
$\hat x= x\cos \theta + y\sin \theta$ and does not depend on the orthogonal 
linear combination $\hat y= -x\sin \theta + y\cos \theta$. 
The BPS equation becomes 
\begin{equation}
{d {\cal W} \over d \hat x} 
=
{\rm e}^{i \theta} 
2{\partial {\cal W} \over \partial z} 
= 
{\rm e}^{i \theta} 
 \Omega_{-}
\left(
K^{ i j^*} {\partial {\cal W} \over \partial A^i}
 \frac{\partial {\cal W}^*}{ \partial A^{*j}}
\right). 
\end{equation}
By defining a new variable with a constant phase factor, 
$\tilde {\cal W}\equiv {\rm e}^{i \theta}  \Omega_{-}{\cal W}$, we find 
\begin{equation}
{d \tilde{\cal W} \over d \hat x} 
=
K^{ i j^*} {\partial  \tilde{\cal W} \over \partial A^i}
 \frac{\partial  \tilde{\cal W}^*}{ \partial A^{*j}}
. 
\label{eq:straight_line_W}
\end{equation}
Since the right-hand side is a real positive quantity, 
the configuration is always real, if we start from a real value 
for $\tilde {\cal W}$ at some point in $x$. 
Therefore the BPS configuration becomes a straight line in the complex 
plane of superpotential ${\cal W}$ for a one-dimensional configuration 
in base space $z$ such as a wall configuration. 
In this configuration, the anti-BPS equation (\ref{Be1}) 
is also valid 
\begin{equation}
2 
{\partial A^i \over \partial \bar{z}} 
=
{\rm e}^{i \theta} 
{d A^i \over d \hat x} 
=
{\rm e}^{2i \theta} 
2 
{\partial A^i \over \partial {z}} 
= 
{\rm e}^{2i \theta} 
\Omega_{-}K^{ i j^*} \frac{\partial {\cal W}^*}{ \partial A^{*j}}. 
\end{equation}
Therefore the configuration is a ${1 \over 2}$ BPS state. 

In order to exploit this behavior of the ${1 \over 2}$ BPS configuration, 
it is useful to use the superpotential ${\cal W}$ as one of the 
chiral scalar superfield. 
This is always achieved by a holomorphic field redefinition. 
{}If we take superpotential as a chiral scalar superfield 
for a nonlinear sigma model, 
the condition of SUSY vacuum reduces to singularities 
of the K\"ahler metric, since there are no stationary points 
of superpotential. 
Even in discussing junction configuration, this choice of superpotential as 
a chiral scalar superfield is still quite useful. 
This is because junction configuration reduces to a domain wall 
asymptotically along each individual wall. 
Therefore an infinite circle around the junction 
is mapped to 
a closed circuit of straight line segments connecting adjacent vacua 
if it is mapped into the complex plane of superpotential ${\cal W}$. 
We shall use this representation of asymptotic behavior of junction 
frequently in later sections.

\vspace{5mm}

\subsection{Holomorphically factorizable nonlinear sigma models 
}
\label{sc:LSM}
Let us first characterize the class of nonlinear sigma models 
which 
can be mapped into linear sigma  models by holomorphic 
field redefinitions. 
If a holomorphic field redefinition 
(\ref{eq:transformation_kahler}) can be made 
from scalar fields $A^i$ of a nonlinear sigma 
model to fields $A'^i$ of a linear sigma model with minimal 
kinetic term, 
the K\"ahler metric of the nonlinear sigma model is given by 
\begin{equation}
K_{i j^*} 
= {\partial A'^{k} \over \partial A^{i}} 
 {\partial A'^{*l} \over \partial A^{*j}} 
\delta_{k l} 
= {\partial A'^{k} \over \partial A^{i}} 
 {\partial A'^{*k} \over \partial A^{*j}} .
\label{eq:LSMcondition}
\end{equation}
This form is the condition for the nonlinear sigma model 
which can be 
mapped into linear sigma model by a holomorphic field 
redefinition. 
We shall call this class of nonlinear sigma models as 
holomorphically 
factorizable. 
In the case of nonlinear sigma models with only a single chiral 
scalar superfield, 
the condition (\ref{eq:LSMcondition}) reduces to the 
factorization 
of K\"ahler metric into holomorphic and anti-holomorphic factors 
\begin{equation}
K_{A A^*} 
 = 
\left| {\partial A' \over \partial A} \right|^2 .
\label{eq:LSMcondition2}
\end{equation}

{}For single field models in this class, 
the SUSY condition reduces to 
\begin{equation}
0=
 {\partial A \over \partial A'} 
  {\partial {\cal W} \over \partial A} 
 = 
  {\partial {\cal W} \over \partial A'} .
\end{equation}
If $\partial {\cal W}/\partial A'$ vanishes along a line segment on the 
complex plane of the field $A$, it should vanish everywhere. 
Therefore we can have only discrete SUSY vacua in this case. 

\begin{figure}[t]
\begin{center}
\leavevmode
\epsfxsize=8cm
\epsfysize=5cm
\centerline{\epsfbox{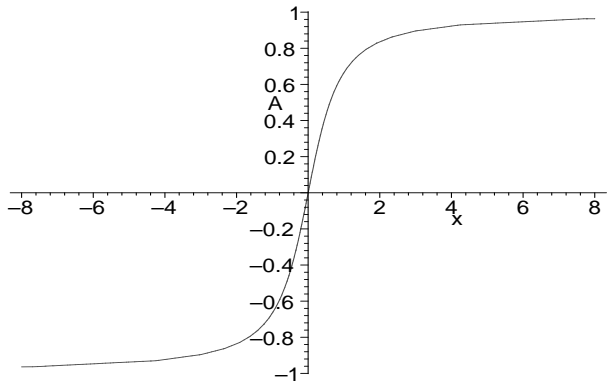}}
\caption{The profile of the wall in Eq.~(\ref{eq:wall1})}
\label{FIG:wall_profile}
\end{center}
\end{figure}

Let us present an example of this class of nonlinear sigma 
models 
to illustrate points raised in the previous section. 
From now on we shall use superpotential as the chiral scalar 
field 
of the nonlinear sigma model. 
Nonlinear sigma models with two isolated singularities within 
the holomorphically factorizble models can be put into the 
following form 
by rescaling and shift of field and coordinates 
\begin{equation}
K_{A A^*} 
 = 
 \left|{1 \over 1-A^2} \right|^2, \qquad 
{\cal W}=A.
\label{eq:NLSM_2sing}
\end{equation}
This model has two SUSY vacua at $A=\pm 1$ where the 
K\"ahler metric 
is singular. 
The BPS equation (\ref{Be2}) for 
a wall connecting the vacuum $A=-1$ at $x=-\infty$ to the 
vacuum 
$A=1$ at $x=\infty$ reduces to 
\begin{equation}
{d A \over d x} 
= K^{A A^*} 
=  \left| 1-A^2 \right|^2 .
\label{eq:wallBPSeq}
\end{equation}
We can easily find the solution 
\begin{equation}
x-x_0 
= {1 \over 4} \left( {2A\over 1-A^2}
+{\rm log}\left({1+A \over 1-A}\right) \right)
\label{eq:wall1}
\end{equation}
where the integration constant $x_0$ corresponds to the 
position of the wall. 
The solution with $x_0=0$ is illustrated in Fig.\ref{FIG:wall_profile}. 
We can also generalize the model to the case of $n$ isolated 
singularities for the K\"ahler metric. 

\begin{figure}[t]
\begin{center}
\leavevmode
\epsfxsize=8cm
\epsfysize=5cm
\centerline{\epsfbox{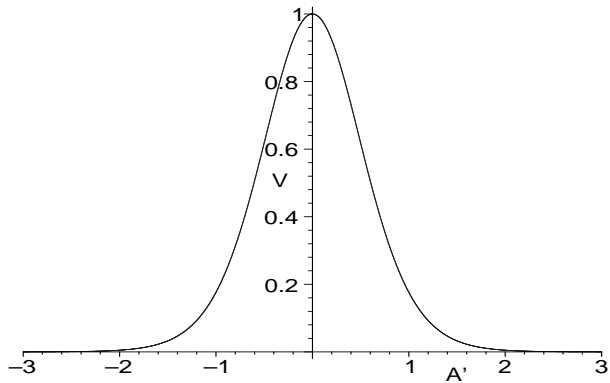}}  
\caption{Scalar potential of the linear sigma model (\ref{eq:ernst_pot})}
\label{FIG:scalar_potential}
\end{center}
\end{figure}

\begin{figure}[t]
\begin{center}
\leavevmode
\begin{eqnarray*}
\begin{array}{cc}
  \epsfxsize=6cm
  \epsfysize=5cm
\epsfbox{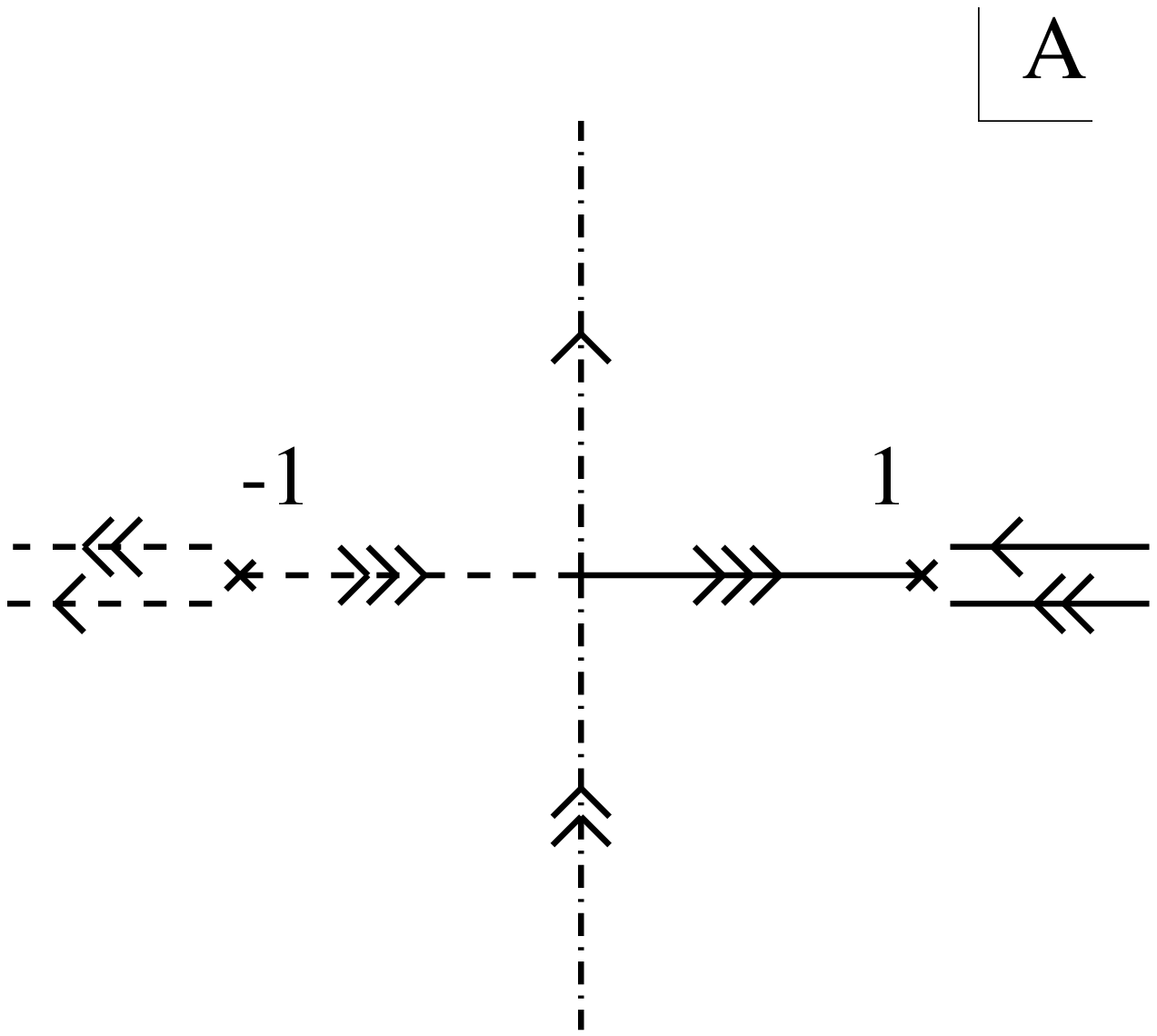} & 
  \epsfxsize=6cm
  \epsfysize=5cm
\hspace*{1cm}
\epsfbox{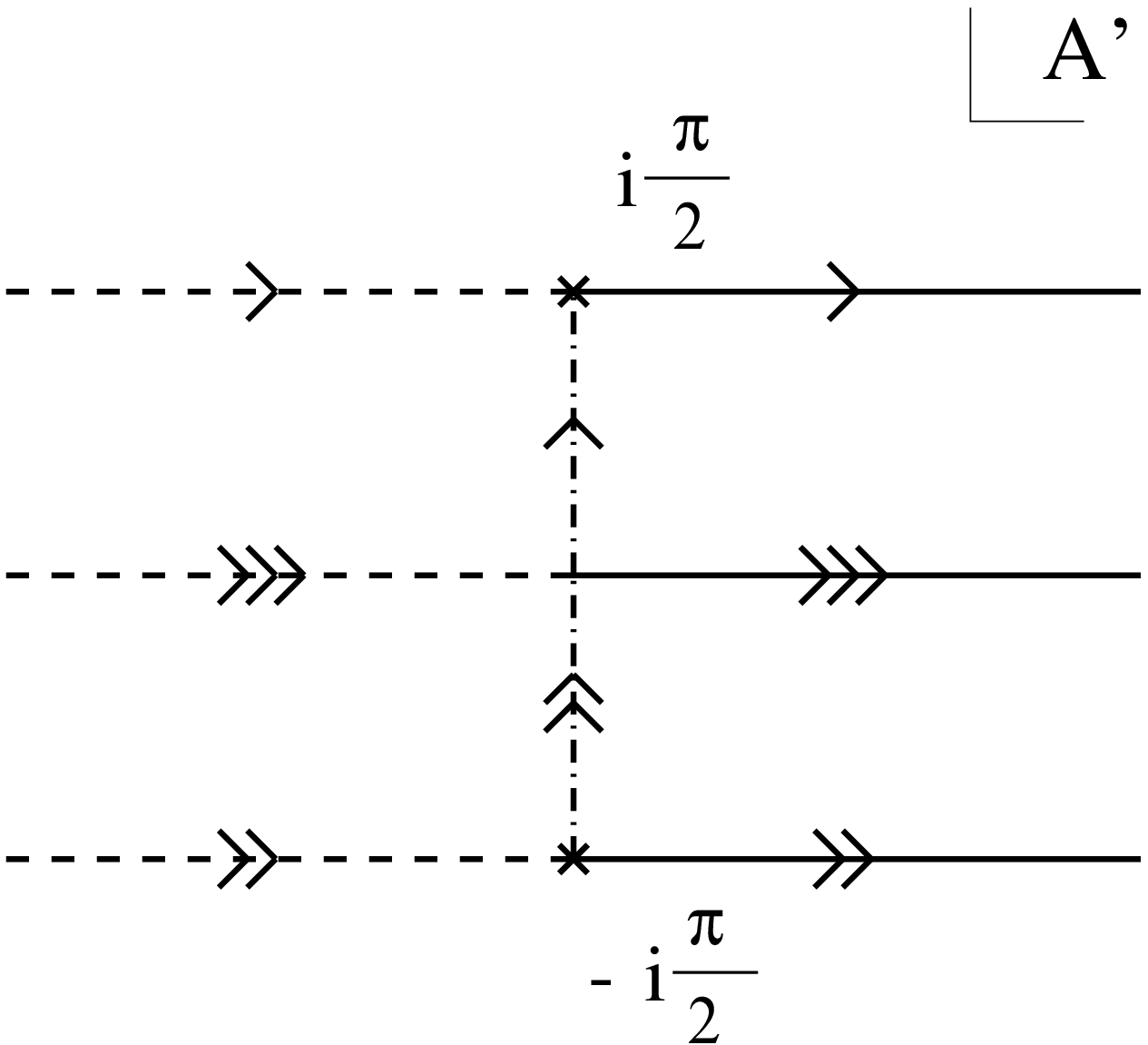} 
\\
\mbox{\footnotesize (a) Nonlinear sigma model} & 
\hspace*{1cm}
\mbox{\footnotesize (b)Linear sigma model}
\end{array} 
\end{eqnarray*} 
\caption{Field space for linear and nonlinear sigma model. 
Solid, dashed, dotted lines with arrows in the $A'$ plane are 
mapped to those corresponding lines in the $A$ plane. }
\label{MODULI-3}
\end{center}
\end{figure}

It is instrutive to map the above model into a linear sigma 
model 
by a holomorphic field redefinition from $A$ to $A'$ 
\begin{equation}
{dA' \over dA} = {1 \over 1-A^2}. 
\end{equation}
The scalar field $A$ of the linear sigma model is given by 
\begin{equation}
A' = {\rm arctanh} A. 
\end{equation}
The bosonic part of the linear sigma model equivalent to the 
nonlinear sigma 
model given in Eq.(\ref{eq:NLSM_2sing}) reads 
\begin{equation}
{\cal L}_{\rm LSM}
=
-\partial_\mu A'^* \partial^\mu A' 
-V(A', A'^*), 
\qquad V(A', A'^*)={1 \over |\cosh^2 A'|^2} .
\label{eq:ernst_pot}
\end{equation}
The SUSY vacua $A=-1$ and $A=1$ of the nonlinear sigma 
model are mapped into 
$A'=-\infty$ and $A'=\infty$ respectively. 
The scalar potential is plotted as a function of $A'$ along the real axis in 
Fig.\ref{FIG:scalar_potential}. 
We see that the scalar potential $V(A', A'^*)$ of the linear sigma model 
vanishes only asymptotically at $A'=\pm \infty$. 
One usually regards these vacua at infinity as the runaway 
vacua and discards them.
The nonlinear sigma model in this example shows an 
advantage of 
revealing these vacua as legitimate SUSY vacua and 
moreover allowing the 
BPS wall solution connecting these two vacua. 
This particular model (\ref{eq:ernst_pot}) was used to discuss the quantum 
tunneling problem \cite{Miller} 
and properties of the "vacuumless" model similar to this one 
have been studied \cite{CVB}. 
The wall solution corresponds to the zero energy limit of the tunneling 
amplitude. 
There is a singularity of the scalar potential at 
$A'=i \left({1 \over 2} + n\right) \pi$ with an integer $n$. 
As shown in Fig.\ref{MODULI-3}, 
 the singularities of K\"ahler 
metric at 
$A=\pm 1$ is mapped to infinity $A'=\infty $ in $A'$ 
plane, and that 
the singularity of the scalar potential at 
$A'=i \left({1 \over 2} + n\right) \pi$ is mapped to 
$A= \infty $ 
in $A$ plane. 
The variable in linear sigma model now becomes a periodic 
variable. 
This appearance of periodic variable carries an interesting 
phenomenon of 
possible winding number which is also noted in \cite{HLS}, \cite{MSSS2}. 
Another interesting example of holomorphically factorizable model 
is given in appendix \ref{NLSM_phys_difference} to illustrate physical 
difference of the K\"ahler metric in discussing nonperturbative 
dynamics of ${\cal N}=1$ SUSY gauge theories. 

\vspace{5mm}

\section{Nonlinear sigma models with exact junction 
solutions}
\label{sc:NLSM_junction}
\subsection{A linear sigma model with $Z_3$ BPS junction
}
\label{sc:LSM_junction}
In this section we study BPS junction configuration in 
nonlinear sigma models. 
Let us first review the exact BPS junction solution obtained in 
Refs.\cite{OINS}, \cite{INOS}. 
The model is a toy model for the low-energy effective theory 
of the 
${\cal N}=2$ $SU(2)$ gauge theory with a single flavor 
\cite{SeibergWitten}, \cite{IntriligatorSeiberg2}. 
It has $U(1)\times U(1)$ gauge symmetry and chiral scalar 
multiplets for ``monopole'' ${\cal M}$, ``anti-monopole'' 
$\bar {\cal M}$, 
``dyon'' ${\cal D}$, ``anti-dyon'' $\bar{\cal D}$ 
and ``quark'' ${\cal Q}$, ``anti-quark'' $\bar{\cal Q}$ 
besides the ``moduli'' $\hat T$. 
Moreover the exact solution was obtained when the model 
possesses the 
$Z_3$ symmetry. 
It has also been recognized that the same exact solution can 
be obtained 
for the model with only chiral scalar superfields by eliminating 
the gauge 
interactions, identifying ${\cal M}=\bar {\cal M}, \ 
{\cal D}=\bar{\cal D}, \ 
{\cal Q}=\bar{\cal Q}$ and adjusting the parameters 
appropriately \cite{INOS}. 
Therefore we can take this linear sigma model and ask if the 
BPS junction 
solution can be generalized to other models and/or other 
symmetries. 
There are four chiral scalar superfields ${\cal M}_i, i=1,2,3$ and $\hat T$ 
with the 
minimal kinetic term. 
The linear sigma model with $Z_3$ symmetric junction 
solution can be 
reduced to the following form of superpotential 
after shift and rescaling of fields 
\begin{equation}
{\cal W}= {1 \over 2}\left[ \sqrt{3} \hat T - 
\sum_{j=1}^3 \left(\sqrt{3}\hat T-{\rm e}^{i{2\pi \over 3}j}\right) 
{\cal M}_j^2 \right]. 
\label{eq:Z3_superpotential}
\end{equation}

\begin{figure}[t]
\begin{center}
\leavevmode
\begin{eqnarray*}
\begin{array}{cc}
  \epsfxsize=6cm
  \epsfysize=6cm
\epsfbox{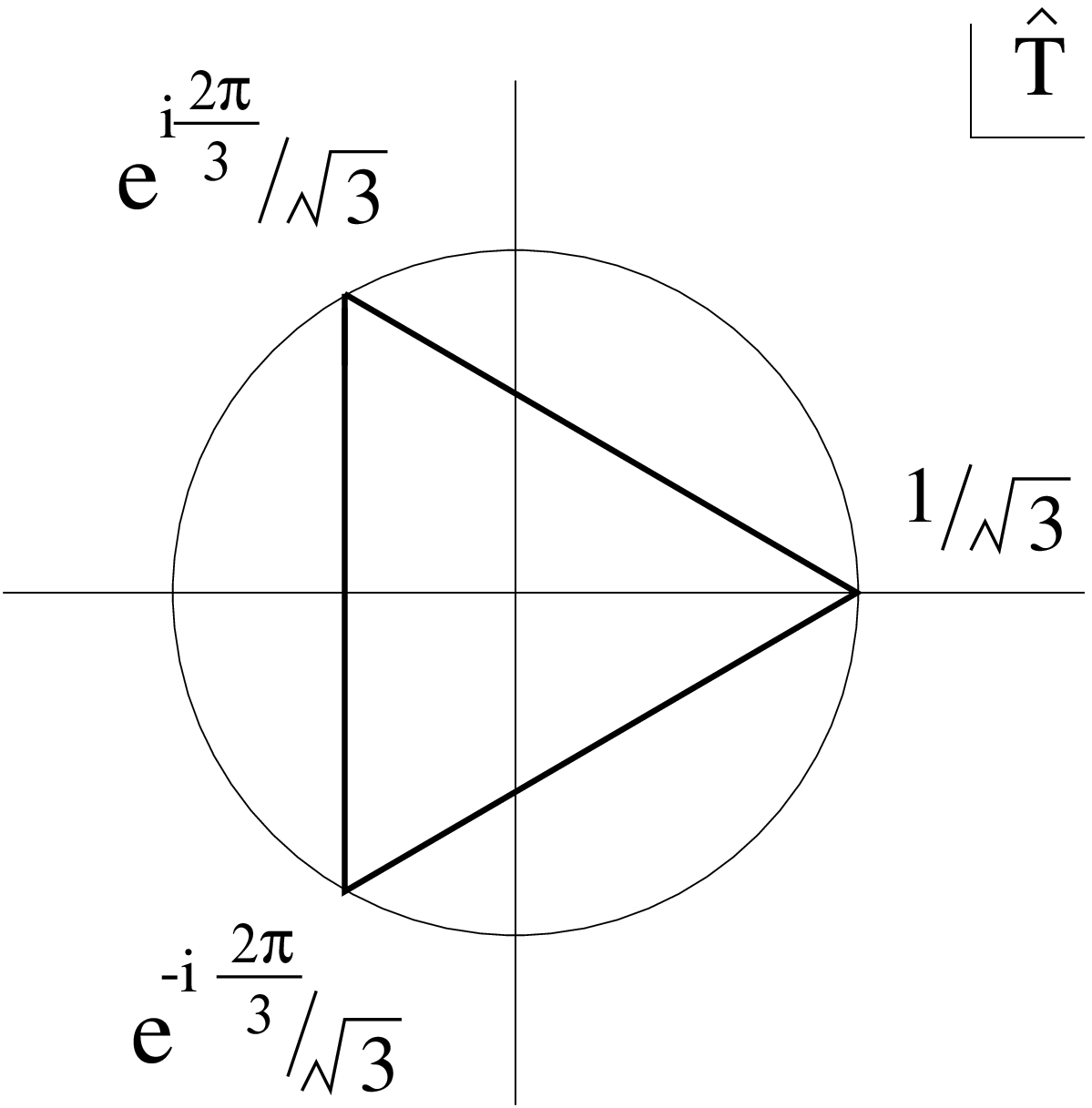} & 
  \epsfxsize=6cm
  \epsfysize=6cm
\hspace*{1cm}
\epsfbox{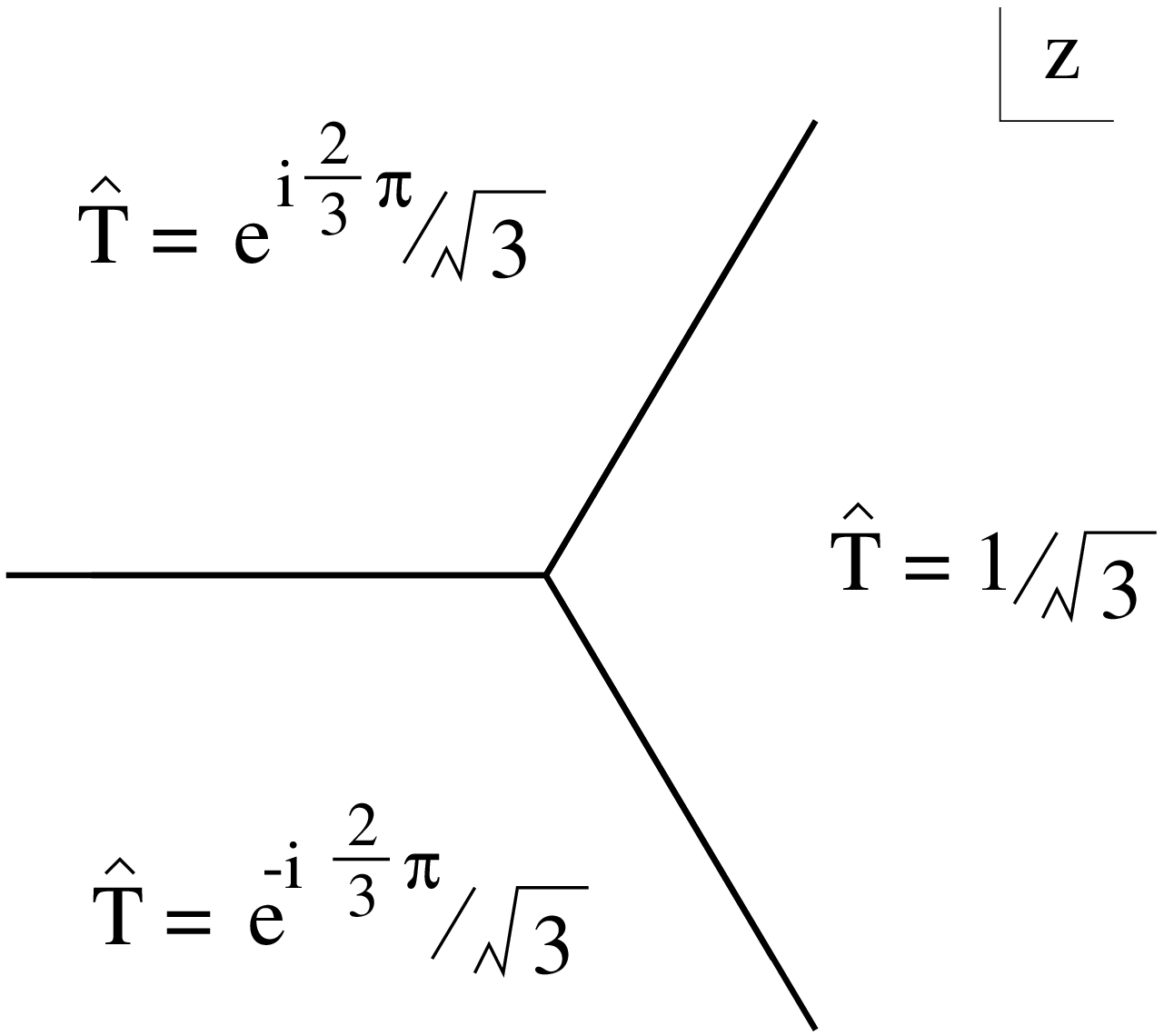} 
\\
\mbox{\footnotesize (a) Vacua of $Z_3$ linear sigma model} & 
\hspace*{1cm}
\mbox{\footnotesize (b) $Z_3$ junction in the base space}
\end{array} 
\end{eqnarray*} 
\caption{$Z_3$ symmetric linear sigma model. 
 }
\label{Z3_linear_model}
\end{center}
\end{figure}

There are three isolated SUSY vacua as depicted in Fig.~\ref{Z3_linear_model} 
\begin{equation}
\hat T={1 \over \sqrt{3}}{\rm e}^{i{2\pi \over 3}j}, 
\quad {\cal M}_j=\pm 1, \quad {\cal 
M}_k=0, \; 
k\not=j, \qquad j=1, 2, 3. 
\label{eq:Z3vacLSM}
\end{equation}
The ${1 \over 4}$ BPS equation (\ref{Be2}) is given by 
\begin{equation}
2 
{\partial {\cal M}_j \over \partial z} 
=  \left( {\rm e}^{-i{2\pi \over 3}j} - \sqrt{3}\hat T^*\right) {\cal M}_j^*, 
\end{equation}
\begin{equation}
2 
{\partial \hat T \over \partial z} 
= {\sqrt{3} \over 2}\left( 1 - \sum_{j=1}^3{\cal M}_j^{*2}\right)
\label{eq:T_BPS_Z3_LSM}
\end{equation}
where we have set the constant phase factor $\Omega_-=1$ 
in order to orient 
the wall separating the vacuum 
$\hat T={\rm e}^{\pm i{2\pi \over 3}}/\sqrt{3}$ along 
the negative real axis as shown in Fig.~\ref{Z3_linear_model}. 
It is convenient to define the following auxiliary 
 quantities taking real values 
\begin{equation}
f_j \equiv \exp \left({1 \over 2}
\left({\rm e}^{-i{2\pi \over 3}j}z
+{\rm e}^{i{2\pi \over 3}j}z^*\right) \right), 
 \qquad j=1, 2, 3. 
\label{eq:fj_Z3}
\end{equation}
which satisfy the following identity 
\begin{equation}
\prod_{j=1}^3 f_j =1. 
\label{eq:identity_fj_Z3}
\end{equation}
The exact BPS solution for a junction connecting three 
vacua is given by \cite{OINS} 
\begin{equation}
\hat T = {1 \over \sqrt{3}}
{\sum_{j=1}^3 {\rm e}^{i{2\pi \over 3}j}f_j \over \sum_{k=1}^3 
f_k}, 
\qquad 
{\cal M}_j = 
{f_j \over \sum_{k=1}^3 f_k}. 
\label{eq:junction_Z3_LSM}
\end{equation}
We have shown that the configuration (\ref{eq:junction_Z3_LSM}) 
on a circle at infinity $|z|\rightarrow \infty$ reduces to a 
collection of three walls separating three vacuum regions 
which are represented by three straight line segments connecting 
three vacua in Fig.~\ref{Z3_linear_model} \cite{OINS}, \cite{INOS}. 

\subsection{A nonlinear sigma model with 
BPS junction
}
We would like to construct a nonlinear sigma model which has 
an exact 
BPS solution of the junction. 
We observe that the junction solution is a mapping from a 
two-dimensional 
base space $x^1, x^2$ to the complex scalar fields ${\cal 
M}_j, j=1,2,3$ and 
$\hat T$. 
Therefore we can reexpress the complex scalar fields ${\cal 
M}_j, j=1,2,3$ 
in favor of the complex coordinate $z=x^1+ix^2$ and then 
invert the relation 
between $\hat T$ and $z$ to express everything in favor of $\hat T$ 
eventually. 
In particular, we can express the right-hand side of the BPS 
equation for $\hat T$ in Eq.(\ref{eq:T_BPS_Z3_LSM}) 
solely in terms of $\hat T$ by this process. 
The resulting BPS equation can be interpreted as the BPS 
equation 
(\ref{Be2}) in a nonlinear sigma model with appropriate 
superpotential and 
K\"ahler metric. 
In the present case, we can make use of the identity 
(\ref{eq:identity_fj_Z3}) 
and use real quantities $f_1$ and $f_2$ instead of 
coordinates $x^1$ and $x^2$ in the 
intermediate stage. 
Using Eqs.(\ref{eq:T_BPS_Z3_LSM}), (\ref{eq:identity_fj_Z3}), and 
(\ref{eq:junction_Z3_LSM}), we obtain 
\begin{eqnarray}
2 
{\partial \hat T \over \partial z} 
&\!\!\!
=
&\!\!\!
{\sqrt{3} \over 2}\left(1 - \sum_{j=1}^3{\cal M}_j^{*2} \right)
= {\sqrt{3} \over 2}\left( 1 - 
{\sum_{j=1}^3 f_j^2 \over \left(\sum_{k=1}^3 f_k\right)^2} \right)
= \sqrt{3}{f_1 f_2 + f_2 f_3 + f_3 f_1 \over \left(\sum_{k=1}^3 
f_k\right)^2}  
\nonumber \\
&\!\!\!
=
&\!\!\!
 \sqrt{3}{f_1 f_2 + {1 \over f_1} + {1 \over f_2} \over 
\left(f_1 + f_2 +{1 \over f_1 f_2}\right)^2}  
=
 \sqrt{3}{f_1 f_2\left(f_1^2 f_2^2 +  f_1 + f_2\right) \over 
\left(f_1 f_2\left(f_1 + f_2\right) +1 \right)^2}. 
\end{eqnarray}
where we have eliminated $f_3$ by means of the identity 
(\ref{eq:identity_fj_Z3}). 
Similarly we can reexpress the field $\hat T=\hat T_R +i \hat T_I$ as 
\begin{equation}
\sqrt{3} \hat T_R 
= {-{f_1+ f_2 \over 2}  + f_3  
\over \sum_{k=1}^3 f_k}  
= {-{f_1 f_2(f_1+ f_2) \over 2}  + 1   
\over f_1 f_2(f_1 + f_2) +1 }  
\end{equation}
\begin{equation}
\sqrt{3} \hat T_I 
= { {\sqrt3 \over 2} (f_1- f_2) 
\over f_1 + f_2 +{1 \over f_1 f_2}}  
= { {\sqrt3 \over 2} f_1f_2(f_1- f_2) \over f_1f_2(f_1 + f_2) +1 } . 
\end{equation}
Therefore we finally find that 
\begin{equation}
2 
{\partial \hat T \over \partial z} 
= { 1 \over \sqrt{3}} \left(1 - 3 \hat T_R^2 - 3 \hat T_I^2\right) 
= { 1 \over \sqrt{3}} \left(1 - |\sqrt{3}\hat T|^2\right) .
\end{equation}
The resulting equation should be identified with a ${1 \over 4}$ BPS 
equation (\ref{Be2}) for a nonlinear sigma 
model. 
We find that the superpotential is linear in $\hat T$ and the K\"ahler metric is nontrivial 
\begin{equation}
K_{\hat T  \hat T^*}
=  {3 \over 1 -  |\sqrt{3}\hat T|^2 }, 
\qquad {\cal W} =\sqrt{3}\hat T .
\end{equation}
We emphasize that the above choice is not a matter of convenience, 
and that the holomorphy forces us to choose the superpotential as 
the chiral scalar superfield $\hat T$ itself except for the freedom of a possible 
proportionality constant ${\cal W}/\hat T$. 
The K\"ahler potential is given by 
\begin{equation}
K(\hat T, \hat T^*)
= \int^{3|\hat T|^2}{dx \over x}\log{1 \over 1 -  x } .
\end{equation}
{}From the above procedure, we see that the nonlinear sigma 
model with a single chiral scalar superfield 
is unique modulo usual freedom of holomorphic field redefinitions, 
if we require that it allows the exact junction 
(\ref{eq:junction_Z3_LSM}) as the ${1 \over 4}$ BPS solution. 

{}By using a rescaled field $T \equiv \sqrt{3} \hat T$, 
 we obtain the nonlinear sigma model 
\begin{equation}
{\cal L}_{{\rm NLSM}}
=
-{1 \over 1-| T|^2} \partial_\mu  T^* \partial^\mu  T 
-\left(1-| T|^2\right) .
\label{eq:junction_NLSM}
\end{equation}
We observe that the nonlinear sigma model has 
 continuous vacua at $| T|=1$, corresponding to the 
continuous 
family of singularities of the K\"ahler metric. 
The moduli space has a topology of a circle. 

\begin{figure}[t]
\begin{center}
\leavevmode
  \epsfxsize=6cm
  \epsfysize=6cm
\epsfbox{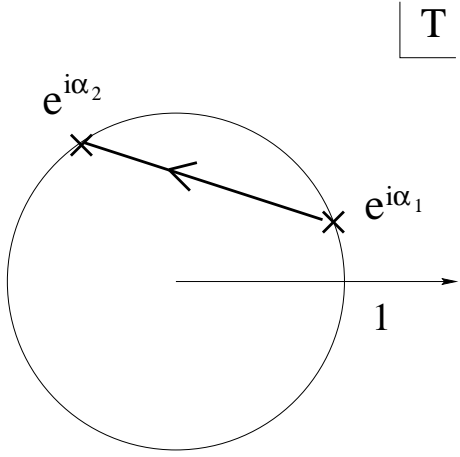} 
\caption{A BPS wall connecting any two vacua}
\label{MODULI-5}
\end{center}
\end{figure}

Now we shall show that there are ${1 \over 2}$ BPS wall 
solutions 
connecting any two arbitrary points in this moduli space. 
Since an orthogonal section of the ${1 \over 2}$ BPS configuration 
should follow a straight line 
trajectory in the 
complex plane of the superpotential, we just need to 
construct a straight line 
connecting two vacua in the moduli space, thanks to our 
choice of the 
superpotential as the field variable. 
Let us define a variable taking a real value $-1 \le \tilde{ T} 
\le 1$ 
along the straight line connecting from 
$  T={\rm e}^{i \alpha_1}$ 
and 
$  T={\rm e}^{i \alpha_2}$ 
as shown in Fig. \ref{MODULI-5} 
\begin{equation}
  T
={{\rm e}^{i \alpha_2} + {\rm e}^{i \alpha_1} \over 2} 
+{{\rm e}^{i \alpha_2} - {\rm e}^{i \alpha_1} \over 2} \tilde{T} .
\end{equation}
Then the ${1 \over 2}$ BPS equations (\ref{Be2}) and 
(\ref{Be1}) 
for the wall reduce to 
\begin{equation}
{d \tilde{T} \over d x} 
= \sin {\alpha_2-\alpha_1 \over 2} 
 \left[ 1-\left(\tilde{T}\right)^2 \right] 
\end{equation}
where we have taken the constant phase factor as 
$\Omega=i{\rm e}^{i {\alpha_2+\alpha_1 \over 2}}$ in order to 
orient the wall along 
the $x=x^1$ direction. 
We can recognize the familiar BPS equation to give the wall 
solution and find 
\begin{equation}
\tilde{ T} 
= \tanh 
 \left( x \sin{\alpha_2-\alpha_1 \over 2} \right)
\end{equation}
\begin{equation}
  T 
= 
{{\rm e}^{i \alpha_2}{\rm e}^{x \sin {\alpha_2-\alpha_1 \over 2}} 
+{\rm e}^{i \alpha_1}{\rm e}^{-x \sin {\alpha_2-\alpha_1 \over 2}} 
\over {\rm e}^{x \sin {\alpha_2-\alpha_1 \over 2}} +
{\rm e}^{-x \sin {\alpha_2-\alpha_1 \over 2}}} .
\label{eq:domainwall}
\end{equation}

\begin{figure}[t]
\begin{center}
\leavevmode
\begin{eqnarray*}
\begin{array}{cc}
  \epsfxsize=6cm
  \epsfysize=6cm
\epsfbox{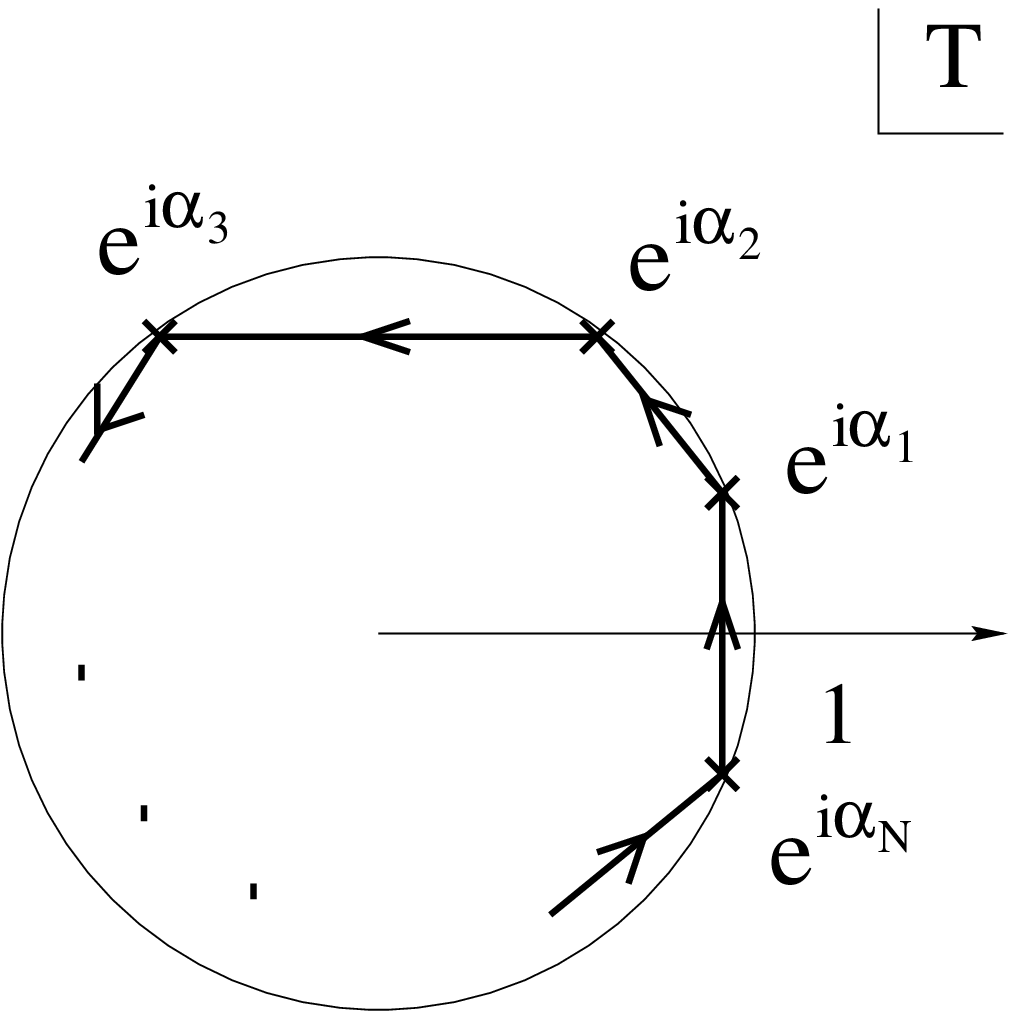} & 
  \epsfxsize=6cm
  \epsfysize=6cm
\hspace*{1cm}
\epsfbox{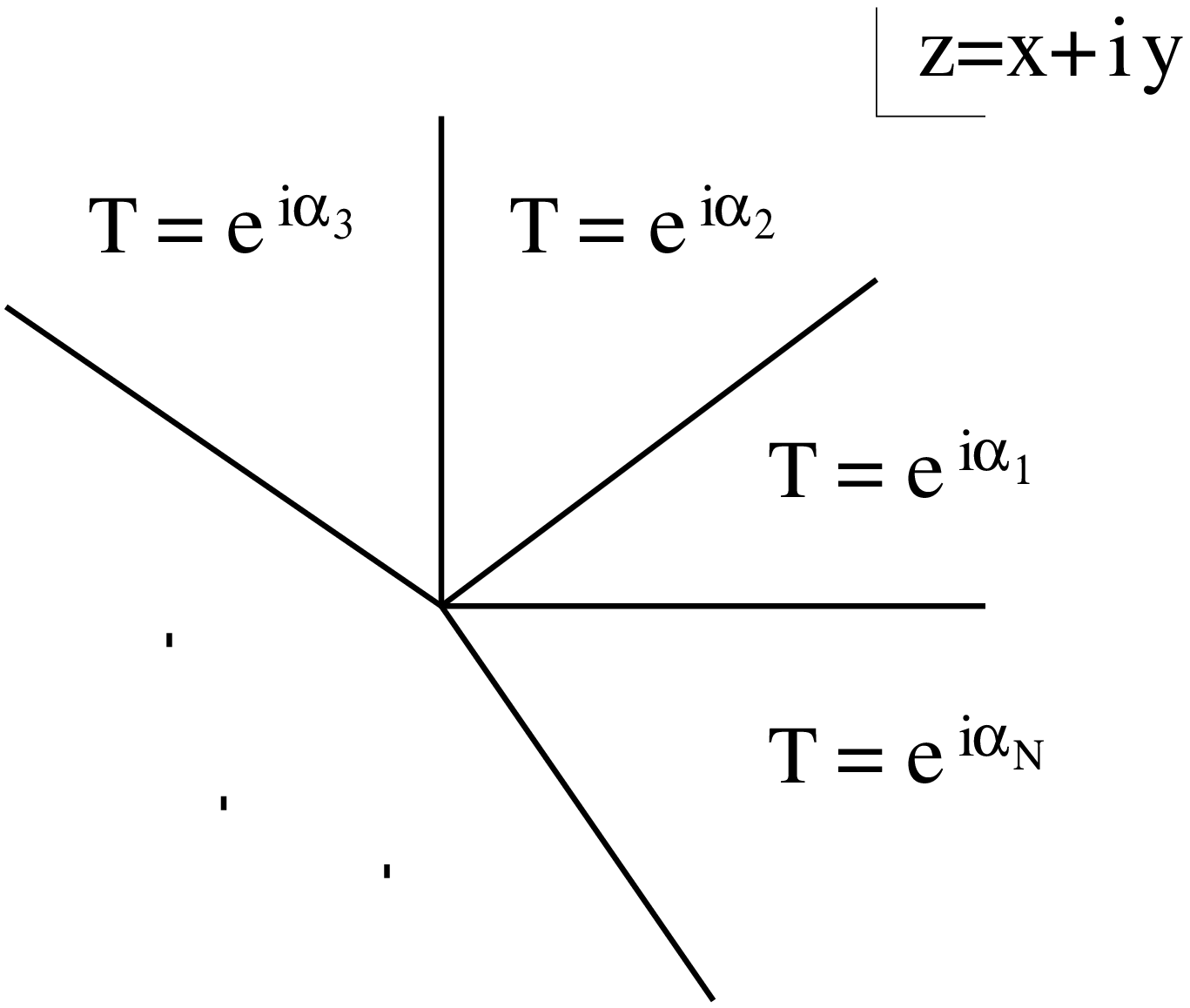} 
\\
\mbox{\footnotesize (a) Moduli space and $N$ vacua for a junction.} & 
\hspace*{1cm}
\mbox{\footnotesize (b) $N$ junction configuration. }
\end{array} 
\end{eqnarray*} 
\caption{Nonlinear sigma model with single chiral scalar superfield 
(\ref{eq:junction_NLSM})}
\label{FIG:n_junction_NLSM}
\end{center}
\end{figure}

More surprisingly, we can show that there exist 
exact ${1 \over 4}$ BPS junction solutions connecting any 
number of 
ordered points ${\rm e}^{i\alpha_j}, j=1, \cdots, N$ in the 
moduli space as shown in Fig.~\ref{FIG:n_junction_NLSM}. 
To obtain the junction solution, let us define the following 
 auxiliary quantities $f_j$ generalized from Eq.(\ref{eq:fj_Z3}) 
\begin{equation}
f_j \equiv \exp \left({1 \over 2}\left({\rm e}^{-i\alpha_j} z
+{\rm e}^{i\alpha_j}z^*\right) \right), 
\qquad j=1, \cdots, N .
\label{eq:fj_general}
\end{equation}
Let us define an auxiliary variable $\eta$ as 
\begin{equation}
\eta \equiv 2\log \left( \sum_{j=1}^N f_j \right) . 
\label{eq:eta}
\end{equation}
Then we obtain 
\begin{equation}
{\partial \eta \over \partial z} =
 {\sum_{k=1}^N {\rm e}^{-i\alpha_k} f_k \over \sum_{j=1}^N f_j}, 
 \qquad 
{\partial^2 \eta \over \partial z \partial z^*} =
{1 \over 2} \left[1 - \left| {\partial \eta \over \partial z} 
\right|^2 
\right], 
\end{equation}
If we take the following Ansatz 
\begin{equation}
  T = {\partial \eta \over \partial z^*} 
= {\sum_{k=1}^N {\rm e}^{i\alpha_k} f_k \over \sum_{j=1}^N f_j}, 
\label{eq:junctionAnsatz}
\end{equation}
we find that the field $T$ satisfies the ${1 \over 4}$ BPS 
equation (\ref{Be2}) for the nonlinear sigma model (\ref{eq:junction_NLSM}) 
with the K\"ahler metric $K_{  T  T^*}=1/(1-|  T|^2)$ and the 
superpotential ${\cal W}=T$
\begin{equation}
2{\partial  T \over \partial z} 
=1 - \left| {\partial \eta \over \partial z} \right|^2 
=1 - \left|  T \right|^2 . 
\label{eq:junction_Be_NLSM}
\end{equation}

\begin{figure}[htbp]
\begin{center}
\leavevmode
\epsfxsize=6cm
\epsfysize=5.5cm
\centerline{\epsfbox{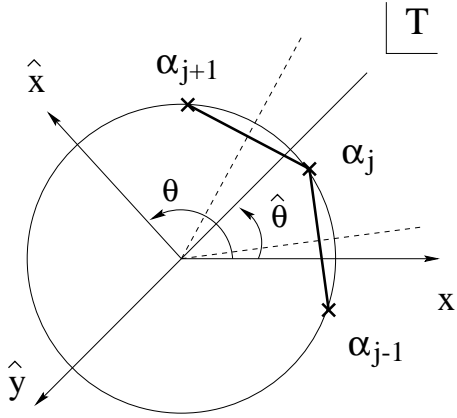}}
\caption{The directions of asymptotic limits of the $N$-juncion 
configuration. 
}
\label{FIG:asymptotic_angles}
\end{center}
\end{figure}

To clarify the physical meaning of this solution, we evaluate the asymptotic 
behavior at infinity. 
Let us define a coordinate system rotated by an angle $\theta$, 
$\hat z \equiv {\rm e}^{-i \theta} z = \hat x + i \hat y$. 
As shown in Fig.~\ref{FIG:asymptotic_angles}, the negative $ \hat y$ axis 
of the rotated coordinates 
is at the angle $ \hat \theta=\theta -\pi/2$. 
Let us first take the asymptotic limit $\hat y \rightarrow - \infty$ 
along a generic direction satisfying 
\begin{equation}
{\alpha_{j-1}+\alpha_j \over 2} < 
\hat \theta < 
{\alpha_{j}+\alpha_{j+1} \over 2} .
\label{eq:generic_limit}
\end{equation}
We find that $f_j$ is dominant compared to all the other $f_k$'s 
and that the $j$-th vacuum is reached asymptotically 
$\hat y \rightarrow - \infty$ 
\begin{equation}
f_j >> f_k, \quad k\not=j, 
\qquad 
 T \rightarrow {\rm e}^{i\alpha_j}. 
\end{equation}
Let us next take the asymptotic limit 
$\hat y \rightarrow - \infty$ along the direction 
$ \hat \theta =(\alpha_{j+1}+\alpha_j)/2$ 
which corresponds to the boundary of the two vacua 
$ T = {\rm e}^{i\alpha_j}$ and  
$ T = {\rm e}^{i\alpha_{j+1}}$, as given by 
the condition (\ref{eq:generic_limit}). 
The auxiliary quantities become for a general $k$ 
\begin{equation}
f_k 
=\exp \left[-\hat y \cos \left(\alpha_k-{\alpha_j+\alpha_{j+1} \over 2}
\right) + \hat x \sin \left(\alpha_k - {\alpha_j+\alpha_{j+1} \over 2}
\right)\right], 
\end{equation}
\begin{eqnarray}
f_j 
&\!\!\!
=
&\!\!\!
\exp \left[-\hat y \cos \left({\alpha_{j+1}-\alpha_{j} \over 2}
\right) - \hat x \sin \left({\alpha_{j+1}-\alpha_{j} \over 2}
\right)\right], 
\nonumber \\
f_{j+1} 
&\!\!\!
=
&\!\!\!
\exp \left[-\hat y \cos \left({\alpha_{j+1}-\alpha_{j} \over 2}
\right) + \hat x \sin \left({\alpha_{j+1}-\alpha_{j} \over 2}
\right)\right] 
\end{eqnarray}
for $j$ and $j+1$ in particular. 
In the limit $\hat y \rightarrow -\infty$, we find 
\begin{equation}
0 \le f_k \ll f_j, f_{j+1}, 
\end{equation}
Therefore we obtain the asymptotic behavior for a wall connecting 
two vacua ${\rm e}^{i\alpha_j}$ to ${\rm e}^{i\alpha_{j+1}}$ 
\begin{equation}
 T \rightarrow 
{{\rm e}^{i \alpha_{j+1}}{\rm e}^{x \sin {\alpha_{j+1}-\alpha_j \over 2}} 
+{\rm e}^{i \alpha_j}{\rm e}^{-x \sin {\alpha_{j+1}-\alpha_j \over 2}} 
\over {\rm e}^{x \sin {\alpha_{j+1}-\alpha_j \over 2}} +
{\rm e}^{-x \sin {\alpha_{j+1}-\alpha_j \over 2}}}, \quad 
 \hat y \rightarrow -\infty,  
\end{equation}
which can be compared with the domain wall solution in 
Eq.(\ref{eq:domainwall}). 
Thus we find that the domain wall 
connecting from the vacuum ${\rm e}^{i\alpha_j}$ to the vacuum 
${\rm e}^{i\alpha_{j+1}}$ 
extends along a direction in the base space $z$ 
which is orthogonal to the line connecting from 
the vacuum ${\rm e}^{i\alpha_j}$ to the vacuum 
${\rm e}^{i\alpha_{j+1}}$ in 
the complex $  T$ plane. 
Moreover the junction solution maps the asymptotic infinity of the 
base space $z$ to a collection of line segments connecting the vacua 
in conformity with the theorem (\ref{eq:straight_line_W}) 
as shown in Fig.~\ref{FIG:n_junction_NLSM}. 

The kinetic term of the nonlinear sigma model is nonnegative 
definite 
only for $| T| \le 1$. 
It is interesting to observe that the moduli space $| T|=1$ of 
the nonlinear 
sigma model forms a natural boundary of the field space 
beyond 
which the kinetic term of the nonlinear sigma model 
is no longer positive definite. 
The curvature of the metric $K_{TT^*}=1/(1-|T|^2)$ is given by 
\begin{equation}
R =2R_{TT^*}\left(1-|T|^2\right) 
=-2R_{TT^*TT^*}\left(1-|T|^2\right)^2={-2 \over 1-|T|^2} .
\end{equation}
Another peculiar feature of the nonlinear sigma model 
(\ref{eq:junction_NLSM}) is that the boundary condition can 
be deformed 
continuously since the moduli space of vacua is continuous. 
The walls and junctions are stable as long as the boundary 
condition is fixed. 
However, the adiabatic change of the boundary condition can 
make the adiabatic 
deformation of the walls and junctions. 

\section{
 Nonlinear sigma models with 
discrete vacua and junction solution 
}
If one wants to prevent the continuous deformation of walls 
and junctions in 
the nonlinear sigma model (\ref{eq:junction_Be_NLSM}), we 
can invent 
another model with discrete vacua. 
We shall work out the model which has the $N$-junction as 
a solution of ${1 \over 4}$ BPS equation. 
Let us introduce the chiral scalar superfields ${\cal M}_j, j=1, 
\cdots, N$ 
besides the chiral scalar superfield $T$. 
We assume the minimal kinetic term for these additional fields 
${\cal M}_j$. 
However, we leave the K\"ahler metric $K_{TT^*}$ of the 
field $T$ to be an arbitray function of $T, T^*$ and 
will determine it by requiring that the model possesses a 
$N$-junction as a solution of the ${1 \over 4}$ BPS 
equation. 
We assume a 
generalization of the superpotential 
in Eq.(\ref{eq:Z3_superpotential})
\begin{equation}
{\cal W}= {1 \over 2}\left[ T - 
\sum_{j=1}^N \left(T-{\rm e}^{i\alpha_j}\right) 
{\cal M}_j^2\right] 
\label{eq:N_superpotential}
\end{equation}
where the parameters $\alpha_j, \; j=1,\cdots,N$ specify the position 
of discrete vacua as we show below, and becomes 
$\alpha_j=2\pi j/N$ for the $Z_N$ symmetric case. 

The SUSY vacuum condition (\ref{eq:SUSYcondition}) 
gives two conditions in this model 
\begin{equation}
 \left({\rm e}^{i\alpha_j}-T\right) {\cal M}_j=0, \qquad 
j=1, \cdots, N, 
\label{eq:mj_susy_cond}
\end{equation}
\begin{equation}
K^{T T^*} {1 \over 2}\left| 1 - 
\sum_{j=1}^N {\cal M}^2_j\right|^2=0. 
\label{eq:t_susy_cond}
\end{equation}
We find that there are two types of the SUSY vacua as in 
Fig.\ref{FIG:n_junction_NLSM}(a): 
\begin{enumerate}
\item 
The $j$-th condition (\ref{eq:mj_susy_cond}) is satisfied 
if the field T takes a particular value $T={\rm e}^{i\alpha_j} $ 
for an integer $j$. 
Then the other conditions (\ref{eq:mj_susy_cond}) imply ${\cal M}_k=0$ 
for $k\not=j$. 
Assuming that the 
 K\"ahler metric is not singular at the point $T={\rm e}^{i\alpha_j} $, 
the second condition (\ref{eq:t_susy_cond}) is only satisfied 
by ${\cal M}_j=\pm 1$. 
\begin{equation}
T={\rm e}^{i\alpha_j}, \quad {\cal M}_j=\pm 1, \quad {\cal 
M}_k=0, \; 
k\not=j, \qquad j=1, \cdots, N. 
\label{eq:ZNvacLSM}
\end{equation}
We will check later that the K\"ahler metric is indeed not singular at this 
point. 
This is the $N$ discrete SUSY vacua given by stationary points of 
the superpotential as shown in Fig.~\ref{FIG:n_junction_NLSM}(a). 
\item 
If $T \not= {\rm e}^{i\alpha_j}, \; j=1, \cdots, N$, 
all the other fields have to vanish ${\cal M}_j=0$ to satisfy 
Eq.(\ref{eq:mj_susy_cond}). 
The other condition (\ref{eq:t_susy_cond}) can only be satisfied 
by a singularity of the K\"ahler metric 
\begin{equation}
K_{TT^*}=\infty. 
\label{eq:kahlersing}
\end{equation}
\end{enumerate}

The ${1 \over 4}$ BPS equations (\ref{Be2}) for ${\cal M}_j$ and $T$ are 
given by 
\begin{equation}
2 
{\partial {\cal M}_j \over \partial z} 
=  \left(  {\rm e}^{-i\alpha_j}-T^*\right) {\cal M}_j^*, 
\label{eq:BPSeq_M}
\end{equation}
\begin{equation}
2 
{\partial T \over \partial z} 
=  K^{TT^*}{1\over 2} \left( 1 - \sum_{j=1}^N{\cal M}_j^{*2}\right) .
\label{eq:T_BPS_ZN_LSM}
\end{equation}
Using the auxiliary quantities $f_j$ defined in Eq.(\ref{eq:fj_general}), 
we assume the following Ansatz for the junction 
\begin{equation}
T = 
{\sum_{j=1}^N {\rm e}^{i\alpha_j}f_j \over \sum_{k=1}^N 
f_k}, 
\qquad 
{\cal M}_j = 
{f_j \over \sum_{k=1}^N f_k}. 
\label{eq:junction_ZN_LSM}
\end{equation}
By using identities
\begin{equation}
{\partial f_j \over \partial z} = 
{1 \over 2} {\rm e}^{-i\alpha_j}f_j, 
\end{equation}
\begin{eqnarray}
{\partial \over \partial z} \left({ f_j \over \sum_{k=1}^N f_k}\right)
&\!\!\!
=
&\!\!\!
{1 \over 2} {{\rm e}^{-i\alpha_j} f_j \over \sum_{k=1}^N f_k}
-{f_j  \over 2} 
{\sum_m {\rm e}^{-i\alpha_m}f_m \over \left(\sum_{k=1}^N f_k\right)^2}
=
 { f_j \over \sum_{k=1}^N f_k}
{1 \over 2} \left( {\rm e}^{-i\alpha_j}-T^* \right)  
\end{eqnarray}
we find that the ${1 \over 4}$ BPS equation for ${\cal M}_j$ 
in Eq.(\ref{eq:BPSeq_M}) is satisfied. 
Then the remaining ${1 \over 4}$ BPS equation 
for $T$ is also satisfied if and only if the 
 K\"ahler metric of the field $T$ is given by 
\begin{eqnarray}
K_{TT^*} 
&\!\!\!
= 
&\!\!\!
{{1\over 2}\left( 1 - \sum_{j=1}^N{\cal M}_j^{*2}\right) \over 
 2 {\partial T \over \partial z}}
= 
{1 - {\sum_{j=1}^N f_j^2 \over \left(\sum_{k=1}^N f_k\right)^2} 
\over 2\left(1 - {|\sum_{l=1}^N {\rm e}^{i\alpha_l} f_l|^2 \over 
\left(\sum_{m=1}^N f_m\right)^2 }\right) }
\nonumber \\
&\!\!\!
=
&\!\!\!
{1 -{\sum_{j=1}^N f_j^2 \over \left(\sum_{k=1}^N f_k\right)^2} 
\over 2\left( 1 - |T|^2 \right) } 
=
{2{\sum_{j<l} f_j f_k \over \left(\sum_{k=1}^N f_k\right)^2} 
\over 2\left( 1 - |T|^2 \right) } 
. 
\label{eq:kahler_ZN_LSM}
\end{eqnarray}
Here we have expressed the metric in terms of 
the auxiliary quantities $f_j$ as an intermediate step. 
The numerator of the right-hand side can be expressed in 
terms of the $T=T_R + iT_I$, since $f_j$ are given in terms of $z$ which 
can be expressed in terms of $T_R$ and $T_I$. 
Therefore we have solved the ${1 \over 4}$ BPS equations and obtained 
the K\"ahler metric of the nonlinear sigma model implicitly through 
Eqs.(\ref{eq:fj_general}), (\ref{eq:junction_ZN_LSM}), 
and (\ref{eq:kahler_ZN_LSM}).
It is gratifying to find that the resulting K\"ahler metric is 
real. 
This is a nontrivial requirement which should be imposed on the 
K\"ahler metric. 
The asymptotic behavior of the solution is precisely the same as in the 
previous section. 
Therefore we have found the ${1 \over 4}$ BPS junction solution connecting 
$N$ discrete vacua. 

We now examine the properties of the K\"ahler metric 
(\ref{eq:kahler_ZN_LSM}). 
The singularity can occur only on a circle $|T|=1$. 
However, the junction solution can never touch the circle as long as 
$z$ is finite. 
The circle can be reached only asymptotically as $|z|\rightarrow \infty$. 
In fact, the analysis in the previous section shows that 
the junction solution approaches asymptotically along a generic direction 
to one of the discrete vacua, say ${\rm e}^{i\alpha_j}$. 
On the other hand, the numerator is such that it also vanishes precisely 
at these vacua and 
the K\"ahler metric becomes finite at the discrete vacua 
\begin{equation}
K_{TT^*} \rightarrow 
{1 \over 2\left(1-\cos(\alpha_j-\alpha_{j-1})\right)}, 
\label{eq:kahler_vacuum}
\end{equation}
if the nearest vacuum to ${\rm e}^{\alpha_j}$ is at 
 ${\rm e}^{\alpha_{j-1}}$. 
If the nearest vacuum is at  ${\rm e}^{\alpha_{j+1}}$ instead, 
we should replace  $\alpha_{j-1}$ by $\alpha_{j+1}$. 
This result shows that the K\"ahler metric is in fact nonsigular as 
we have assumed in analyzing the SUSY vacua. 
The junction  asymptotically along the wall direction 
becomes a wall solution. 
In our solution, we have already shown in the previous section 
that the wall is mapped to a straight line segments 
connecting adjacent vacua in the $T$ plane. 
The K\"ahler metric takes the value (\ref{eq:kahler_vacuum}) 
along the straight line corresponding to the wall connecting the vacua 
$T={\rm e}^{\alpha_{j-1}}$ and $T={\rm e}^{\alpha_{j}}$. 
Let us call $R=\{T(z, z^*) \in {\bf C}, z \in {\bf C}\}$ 
the image of the entire real space $z$ by the map $T(z, z^*)$ 
defined by the junction solution (\ref{eq:junction_ZN_LSM}). 
Summarizing the properties of the K\"ahler metric, we find 
\begin{enumerate}
\item 
The K\"ahler metric is real and positive in $R$. 
\item 
The K\"ahler metric is never singular in $R$. 
\item
The field $T$ can approach asymptotically to the unit circle 
only at the discrete vacuum $T={\rm e}^{i\alpha_j}$ where 
the K\"ahler metric is finite. 
\item
The asymptotic value of the junction solution is mapped to a straight 
line segments connecting these discrete vacua. 
\item
At the origin of the base space, the auxiliary quantity becomes $f_j=1$. 
In the case of $Z_N$ symmetric model, 
the field takes values $T=0, {\cal M}_j=1/N$ and 
the K\"ahler metric takes a value $K_{TT^*} = (N-1)/(2N)$. 
\end{enumerate}
These results suggest that $R$ is a polygon with the $N$ discrete vasua 
as vertices. 
It is interesting to observe that the moduli space is divided into 
several discrete stationary points and 
singular arcs. 
We expect 
that the kinetic term is positive 
definite in $|T| \le 1$, and that the natural domain 
of definition for our nonlinear sigma model is $|T|\le 1$, 
which turn out to be the case in models that we are going to 
analyze more explicitly. 

Let us evaluate the K\"ahler metric as a function of $T$ more explicitly. 
{}For that purpose, we shall first take the $Z_N$ symmetric case. 
The model with junction starts from $N=3$. 
The model with $N=3$ turn out to give a minimal kinetic term 
$K_{TT^*}=1$ and hence 
reduces to our original model in sect.\ref{sc:LSM_junction}. 
The next model is $N=4$. 
We find that the model in fact gives a nonlinear sigma model 
for 
the field $T$ as 
\begin{equation}
K_{TT^*} = { 1 \over 2(1 - |T|^2)} 
\left[{3 \over 4} -{1 \over 2}|T|^2 
-{1 \over 16}\left(T^4 
+T^{*4}\right) 
-{1 \over 8} |T|^4 \right] .
\label{eq:kahler_ZN_T}
\end{equation}

\begin{figure}[t]
\begin{center}
\leavevmode
\begin{eqnarray*}
\begin{array}{cc}
  \epsfxsize=6cm
  \epsfysize=6cm
\epsfbox{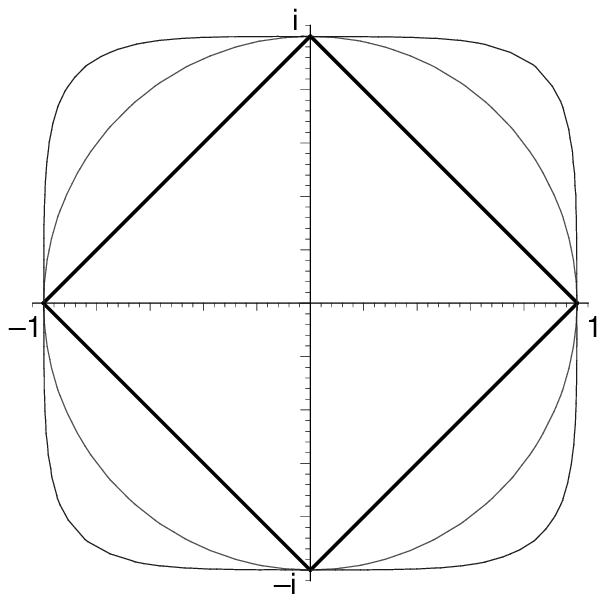} & 
  \epsfxsize=6cm
  \epsfysize=6cm
\hspace*{1cm}
\epsfbox{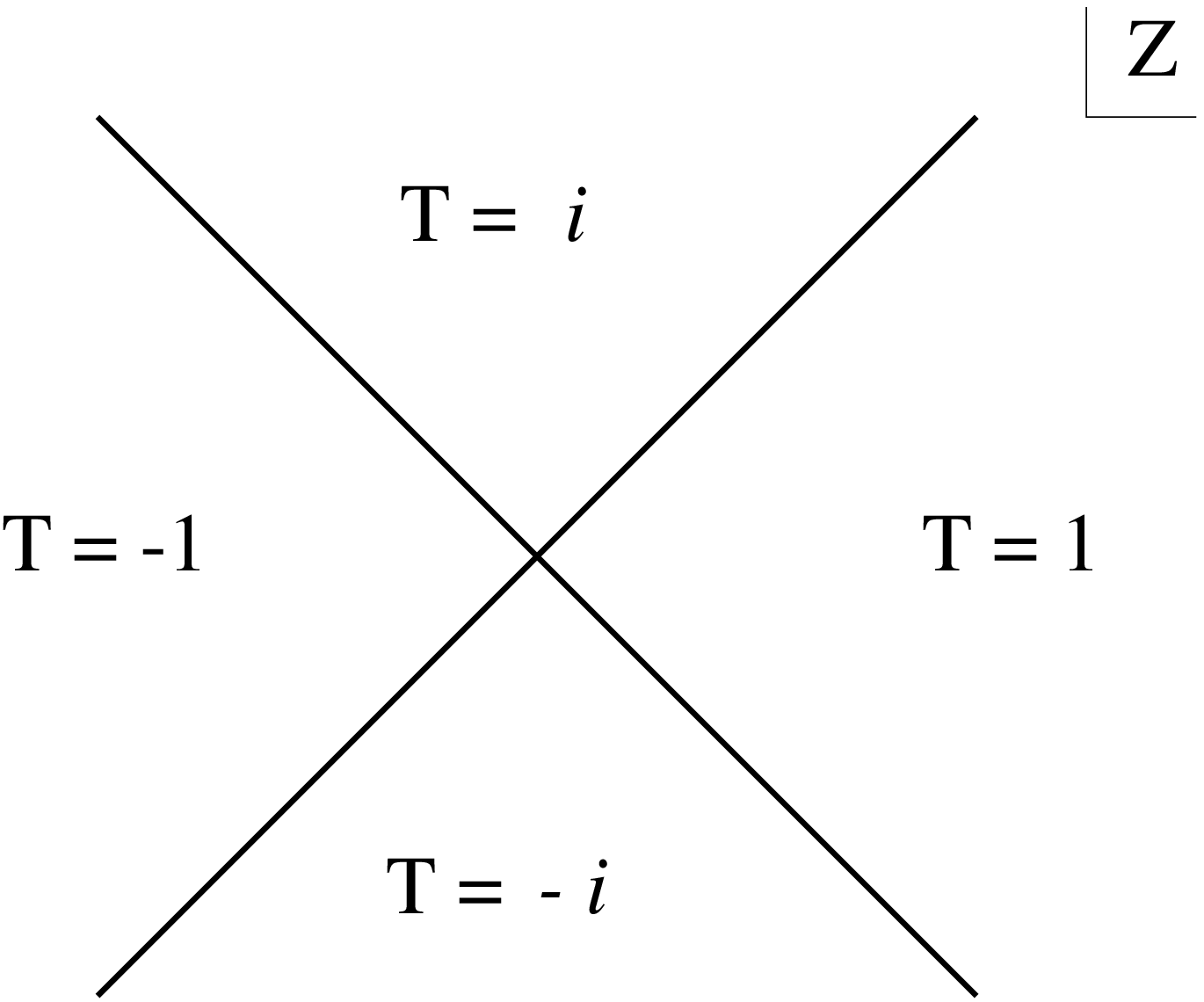} 
\\
\mbox{\footnotesize (a) Vacua of the $Z_4$ symmetric nonlinear sigma model} & 
\hspace*{1cm}
\mbox{\footnotesize (b) $Z_4$ junction}
\end{array} 
\end{eqnarray*} 
\caption{Nonlinea sigma model with $Z_4$ symmetric junction. }
\label{MODULI-8}
\end{center}
\end{figure}
The explicit representation (\ref{eq:kahler_ZN_T}) 
in terms of $T$ shows that 
there is a locus of zeros which touches the unit circle precisely 
at the four discrete points $T={\rm e}^{i{2\pi \over 4}j}, j=1, \cdots, 4$. 
This concrete example shows our general analysis clearly: 
 the K\"ahler metric is singular along 
four arcs separated by discrete stationary points of the superpotential 
$T={\rm e}^{i{2\pi \over 4}j}, j=1, \cdots, 4$ 
where the K\"ahler metric is finite. 
The region between the circle $|T|=1$ and the outer curve touching the circle 
at $T=1, i, -1, -i$ gives the region of negative kinetic term as shown 
in Fig.~\ref{MODULI-8}(a). 
In this model, we have also a continuous moduli space consisting of 
the arcs of singularities of the K\"ahler metric. 
However, even if we attempt to make a wall and/or a junction solution 
connecting these vacua of singularities, we find that 
the BPS equations are not satisfied by Ansatz similar to 
 our previous one in Eqs.(\ref{eq:junctionAnsatz}) or 
(\ref{eq:junction_ZN_LSM}), since 
$T$ now has a nonlinear kinetic term in our model. 
It is possible that there may exist wall or junction solutions 
connecting these singularities of the K\"ahler metric. 
Even if they exist, however, these solutions cannot be adiabatically 
deformed to our solution, since the other fields ${\cal M}$ take 
different values in two types of vacua as shown in Eqs.(\ref{eq:ZNvacLSM})
and (\ref{eq:kahlersing}). 

\begin{figure}[t]
\begin{center}
\leavevmode
\begin{eqnarray*}
\begin{array}{cc}
  \epsfxsize=6cm
  \epsfysize=6cm
\epsfbox{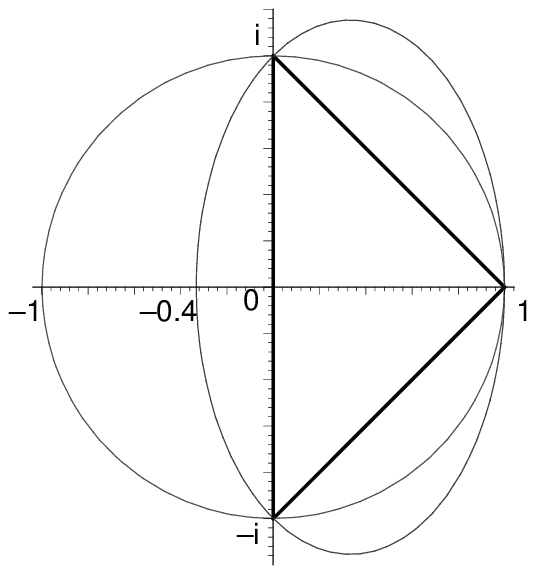} & 
  \epsfxsize=6cm
  \epsfysize=6cm
\hspace*{1cm}
\epsfbox{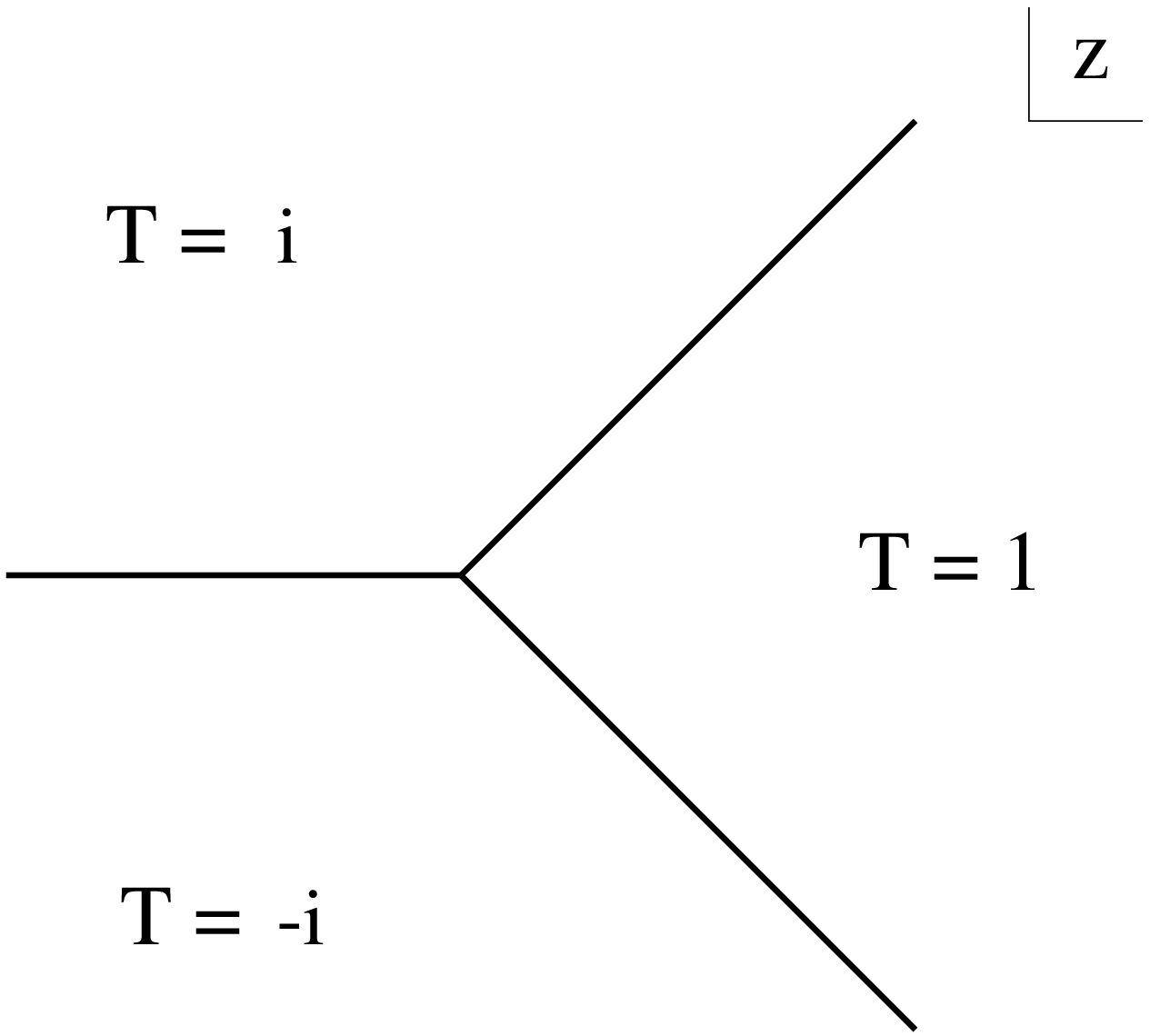} 
\\
\mbox{\footnotesize (a) Vacua at $T=1, i, -i$ 
of a nonlinear sigma model. } & 
\hspace*{1cm}
\mbox{\footnotesize (b) Asymmetric $3$-junction. }
\end{array} 
\end{eqnarray*} 
\caption{Nonlinear sigma model with asymmetric $3$-junction. }
\label{FIG:asymmetric_junction}
\end{center}
\end{figure}

As another example, we shall give a model without the $Z_N$ symmetry. 
Let us take three ``matter'' fields ${\cal M}_j, j=1,2,3$ besides $T$. 
We choose the superpotential (\ref{eq:N_superpotential}) 
with $\alpha_1=\pi/2, \alpha_2=-\pi/2, \alpha_3=0$. 
Then we have 
three discrete vacua at asymmetric points $T=1, i, -i$ on the unit 
circle as shown in Fig.~\ref{FIG:asymmetric_junction}. 
We find that the K\"ahler metric in Eq.(\ref{eq:kahler_ZN_LSM}) 
 becomes in this model 
\begin{equation}
K_{TT^*} = { 1 \over 4(1 - |T|^2)} 
\left(1 +T+T^* - 2|T|^2 
-{T^2+T^{*2} \over 2}\right). 
\label{eq:kahler_asym_T}
\end{equation}
The K\"ahler metric again has a general feature: 
it is finite at the discrete vacua which separate the arcs of singular 
points of the K\"ahler metric as shown in Fig.~\ref{FIG:asymmetric_junction}. 
The kinetic term is negative in the region between 
the circle $|T|=1$ and the ellipse touching the circle 
at three vacua $T=1, i, -i$ as shown in Fig.~\ref{FIG:asymmetric_junction}.

\vspace{1.5cm}

\noindent {\Large \bf Acknowledgments}\\

\noindent 
One of the authors (N.S.) thanks a discussion with Tohru Eguchi, 
Kazuo Fujikawa, and Yoichi Kazama. 
This work is supported in part by Grant-in-Aid 
for Scientific Research from the Japan Ministry 
of Education, Science and Culture  13640269, and those for 
the Priority Area 707.

\renewcommand{\thesubsection}{\thesection.\arabic{subsection}}

\appendix

\section{Example of nontrivial K\"ahler metric and SUSY vacua
} 
\label{NLSM_phys_difference}
\begin{figure}[ht]
\begin{center}
\leavevmode
\begin{eqnarray*}
\begin{array}{cc}
  \epsfxsize=6cm
  \epsfysize=6cm
\epsfbox{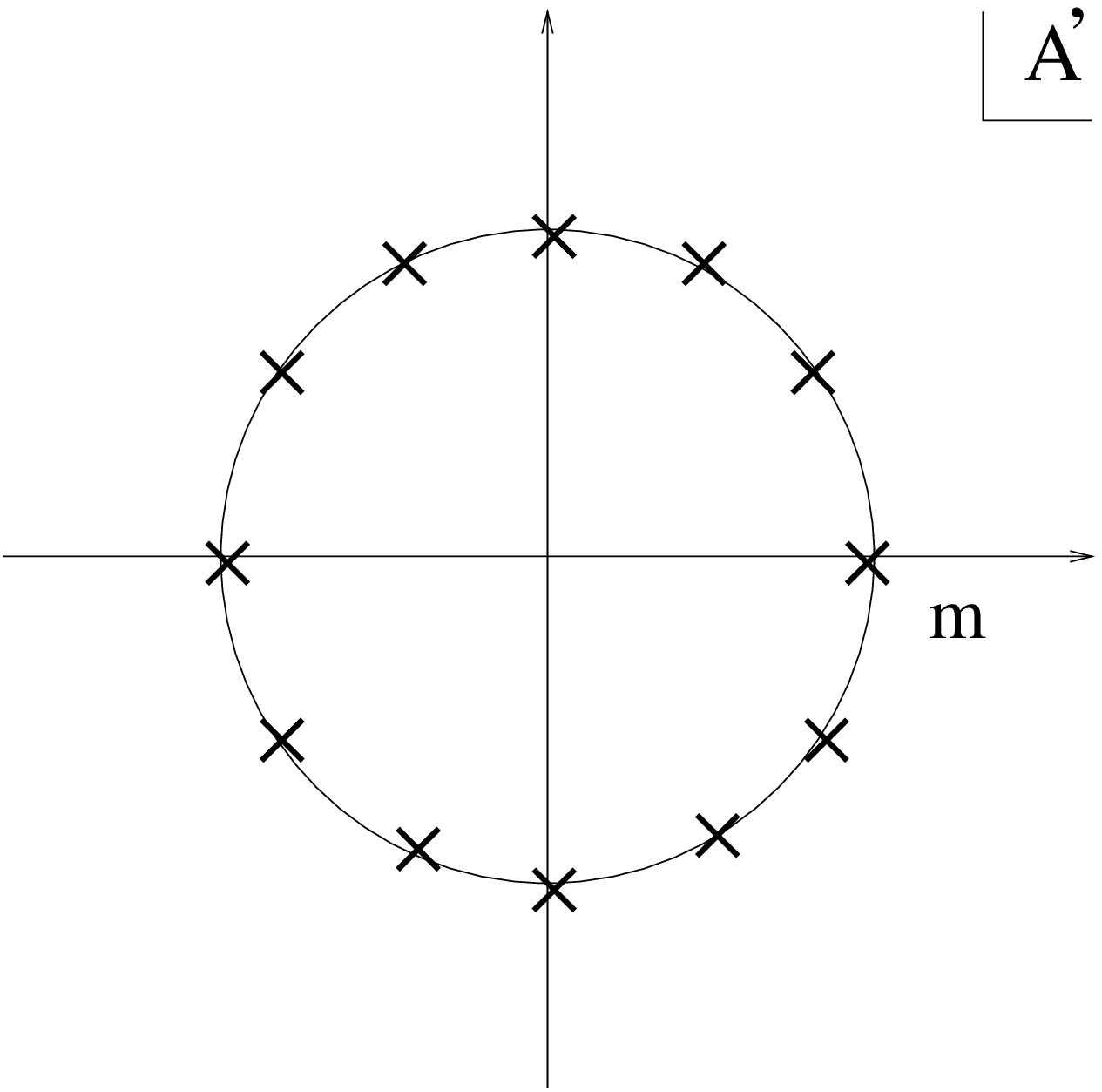} & 
  \epsfxsize=6cm
  \epsfysize=6cm
\hspace*{1cm}
\epsfbox{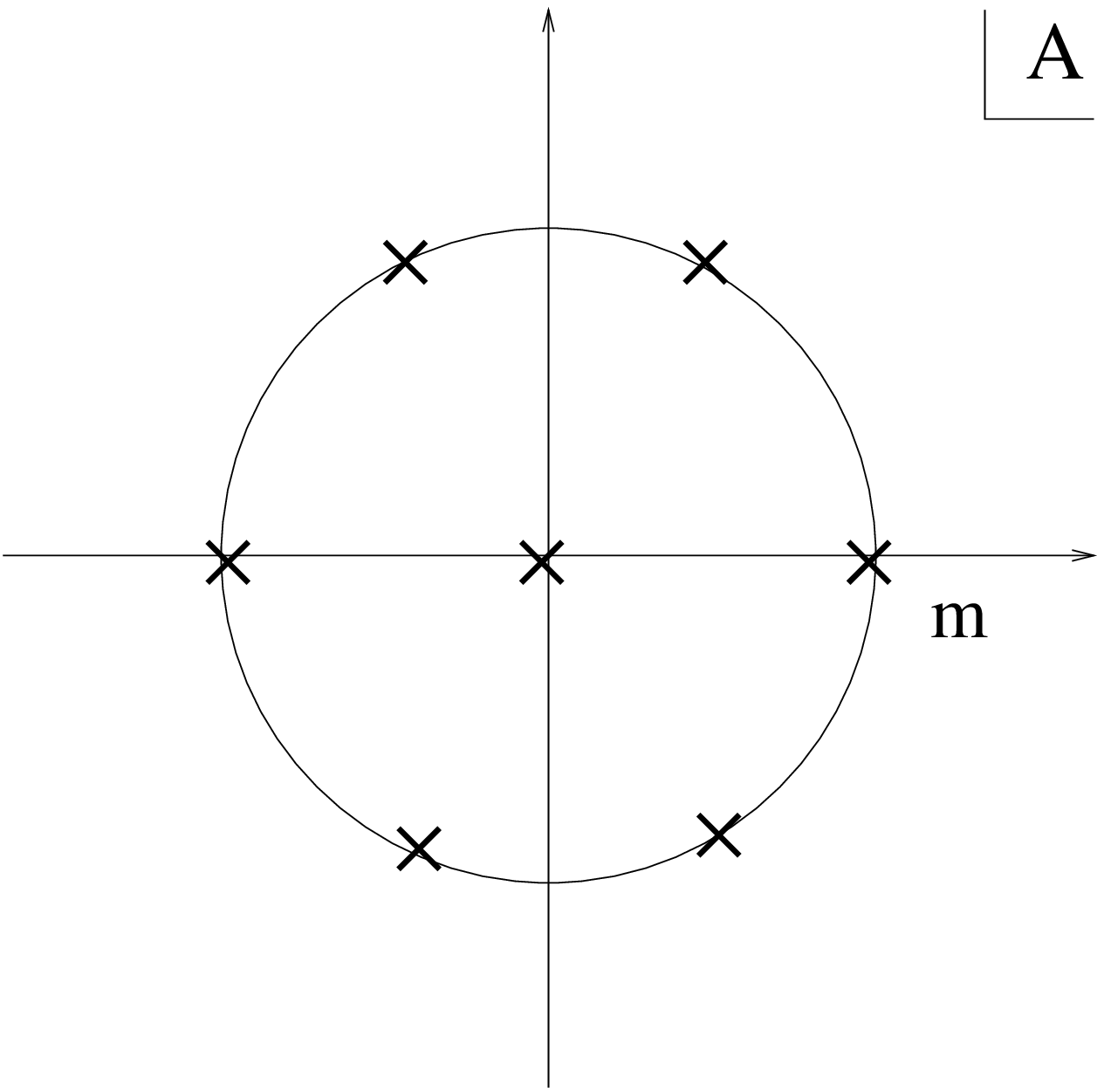} 
\\
\mbox{\footnotesize (a)Vacua in linear sigma model} & 
\hspace*{1cm}
\mbox{\footnotesize (b)Vacua in nonlinear sigma model}
\end{array} 
\end{eqnarray*} 
\caption{}
\label{MODULI-4}
\end{center}
\end{figure}

In this appendix, 
we shall present as another interesting example of holomorphically 
factorizable model, a toy model 
which mimics the difference in physical consequences due to 
the choice of the K\"ahler metric in discussing the nonperturbative 
dynamics of ${\cal N}=1$ SUSY theories. 
The $SU(N_c)$ SUSY gauge theories is known to have 
gaugino condensation 
with $N_c$ distinct chirally asymmetric vacua. 
This gaugino condensation induces a nonperturbative 
superpotential. 
There have been a lot of discussion about the wall solutions 
connecting 
these vacua. 
We shall denote the moduli field by a chiral scalar superfield 
$\Phi'$ corresponding to a color singlet field. 
Let us take the case $N_c=2n$ for simplicity and mimic 
the $2n$ distinct vacua as stationary points at 
$A'={m} {\rm e}^{i {2\pi \over 2n}k}, \; 
k=0, 1, \cdots, 2n-1$
of the superpotential 
\begin{equation}
{\cal W}^{(1)}_{\rm np}(\Phi')
={m^2 }\Phi'
\left(1-{1 \over 2n+1}\left({\Phi' \over m}\right)^{2n}\right). 
\label{eq:np_superpotential}
\end{equation} 
The ${\cal N}=1$ SUSY is not powerful enough to determine the 
kinetic term 
of the ``moduli'' field $\Phi'$. 
Usually one assumes that $\Phi'$ has a minimal kinetic term. 
Then the bosonic part of the Lagrangian with the minimal 
kinetic term and 
with the nonperturbative superpotential 
(\ref{eq:np_superpotential}) 
is given by 
\begin{equation}
{\cal L}^{(1)}
=
-\partial_\mu A'^* \partial^\mu A' 
-\left|{d{\cal W}^{(1)}_{\rm np} \over d A'}\right|^2 
=
-\partial_\mu A'^* \partial^\mu A' 
-\left|m^2 \left(1-\left({A' \over m}\right)^{2n}\right)\right|^2 .
\label{eq:minimal_nonperturbative}
\end{equation}
We shall refer this model as model one. 
The vacua of the model one in the complex $A'$ plane are illustrated 
in Fig.\ref{MODULI-4}(a). 
Numerical evidence showed that there are wall solutions 
connecting 
the adjacent vacua at least \cite{DW}. 

Another school of thought for the kinetic term of the moduli field assumes 
a nonlinear kinetic term. 
This is suggested by an idea that the moduli field is a color singlet 
composed of a bilinear of colored field. 
To illustrate the effect of nonlinear kinetic term, let us suppose that 
the moduli field $\Phi$ in this second model corresponds to the square 
 of a chiral scalar superfield $\Phi'$ with a minimal kinetic term 
\begin{equation}
\Phi
=
\Phi'^2/
m
\end{equation}
where a parameter $m$ of mass dimension one is used to 
make the dimension 
of scalar field canonical. 
We shall also assume that 
the nonperturbative superpotential has $n=N_c/2$ 
distinct 
stationary points at $A={m } {\rm e}^{i {2\pi \over n}k}, \; 
k=0, 1, \cdots, n-1$ and becomes 
\begin{equation}
{\cal W}^{(2)}_{\rm np}(A)
={m^2 }A
\left(1-{1 \over n+1}\left({A \over m}\right)^{n}\right)
\label{eq:np_superpotential2}
\end{equation} 
in conformity with the fact that the two stationary points 
in $A'$ are mapped into one point in $A$. 

The bosonic part of the Lagrangian becomes in this case 
\begin{eqnarray}
{\cal L}^{(2)}
&\!\!\!
=
&\!\!\!
-K_{A A^*}
\partial_\mu A^* \partial^\mu A 
-K^{A A^*}
\left|{d{\cal W}^{(2)}_{\rm np} \over d A}\right|^2 
\nonumber \\
&\!\!\!
=
&\!\!\!
-\left|{1 \over 2}\sqrt{m \over A}\right|^2\partial_\mu A^* 
\partial^\mu A 
-\left|2\sqrt{A \over m}\right|^2\left|{m^2 }
\left(1-\left({A \over m}\right)^{n}\right)\right|^2 .
\end{eqnarray}
We shall refer this model as two. 
In this case of the nonminimal kinetic term, we have an 
additional 
SUSY vacuum at the singularity of the K\"ahler metric at 
$A=0$ 
besides the already existing $N_c$ SUSY 
vacua at the stationary point of the superpotential. 
This additional SUSY vacuum mimics the so-called chirally 
symmetric vacuum. 
The physical difference of two models of nonperturbative 
dynamimcs 
originates from the different assumption 
on the K\"ahler potential which is not well-constrained in 
the ${\cal N}=1$ 
SUSY field theories.


\end{document}